\documentclass[useAMS,usenatbib]{mn2e} %ASCII encoding required
\usepackage{amsmath}
\usepackage{graphicx}
\usepackage[T1]{fontenc}
\usepackage{epstopdf}
\title[]{Evolution of dust and molecular hydrogen in the Magellanic System}
\author[C. Yozin and K. Bekki]{C. Yozin\thanks{{\bf E-mail} 21101348@student.uwa.edu.au; kenji.bekki@uwa.edu.au} and K. Bekki\footnotemark[1]
\\\\
ICRAR, M468, The University of Western Australia, 
35 Stirling Highway, Crawley
Western Australia, 6009, Australia}

\begin{document}

\date{Accepted 2014 June 6. Received 2014 May 21; in original form 2013 October 1}

\pagerange{\pageref{firstpage}--\pageref{lastpage}} \pubyear{2013}

\maketitle

\label{firstpage}

\newcommand{\htw}{H$_{\rm 2}$}
\newcommand{\hi}{H~{\sc i}}
\newcommand{\hii}{H$_{\rm II}$}
\newcommand{\ha}{H$\alpha$}
\newcommand{\ex}[2]{#1$\times$10$^{\rm #2}$}
\newcommand{\md}{M$_{\odot}$}
\newcommand{\ao}{A$_{\rm O}$}
\newcommand{\zd}{Z$_{\odot}$}
\newcommand{\mfh}{f$_{\rm H_{\rm 2}}$}
\newcommand{\yr}{yr$^{\rm -1}$}
\newcommand{\cms}{cm$^{\rm -2}$}
\newcommand{\km}{km$s^{\rm -1}$}
\newcommand{\feh}{[Fe\H]}
\newcommand{\gdrn}{GDR$_{\rm NORM}$}
\newcommand{\fgas}{f$_{\rm gas}$}
\newcommand{\fpah}{f$_{\rm pah}$}
\newcommand{\rhoth}{$\rho_{\rm th}$}
\newcommand{\kpc}{$kpc^{\rm -2}$}
\newcommand{\degsq}{deg$^{\rm -2}$}
\newcommand{\tacc}{$\tau_{\rm acc}$}
\newcommand{\tacco}{$\tau_{\rm acc,0}$}
\newcommand{\tdest}{$\tau_{\rm dest}$}
\newcommand{\tdesto}{$\tau_{\rm dest,0}$}
\newcommand{\cmc}{cm$^{\rm -3}$}
\newcommand{\tsn}{$\tau_{\rm sn}$}
\newcommand{\asym}[1]{A$_{\rm #1}$}
\newcommand{\rh}{$M_{\rm h}/M_{\rm d}$}
\newcommand{\rv}{R$_{\rm V}$}
\newcommand{\xco}{X$_{\rm CO}$}
\newcommand{\rdm}{R$_{\rm d:m}$}
\newcommand{\msmc}{M$_{\rm SMC}$}
\newcommand{\mlmc}{M$_{\rm LMC}$}

%%%%%%%%%%%%%%%%%%%%%%%%%%%%%%%%%%%%%%%%%%%%%%%%%%%%%%%%%%%%%%%%%%%%%%%%%%%%%%%
\begin{abstract}

We investigate the evolution of the interstellar medium (ISM) in self-consistent, chemodynamical simulations of the Magellanic Clouds (MCs) during their recent ($z<0.3$) past. An explicit modelling of dust and molecular hydrogen lifecycles enables us to compare our models against the observed properties of the ISM, including elemental depletion from the gas-phase. Combining this model with a tidal-dominated paradigm for the formation for the Magellanic Stream and Bridge, we reproduce the age-metallicity relations, long gas depletion timescales, and presently observed dust and molecular hydrogen masses of the MCs to within their respective uncertainties. We find that these models' enrichment depends sensitively on the processing of dust within the ISM and the dynamical influence of external tides/stellar bars. The ratio of characteristic dust destruction timescales in our SMC and LMC models, a governing parameter of our models' evolution, is consistent with estimates based on observed supernova (SN) rates. Our reference MC models tend to exhibit the disputed universal dust-to-metal ratio, which we argue stems from the adoption of high SNe II condensation efficiencies. Our models are the first to reproduce the one-tenth solar metallicity of the Stream/Leading Arm following tidal stripping of the SMC; the hypothesis that the LMC contributes a metal-rich filament to the Stream, as implied by recent kinematic and abundance analyses, is also appraised in this study.

\end{abstract}

\begin{keywords}
galaxies: interactions -- galaxies: evolution -- galaxies:ISM -- galaxies: Magellanic Clouds
\end{keywords}

%%%%%%%%%%%%%%%%%%%%%%%%%%%%%%%%%%%%%%%%%%%%%%%%%%%%%%%%%%%%%%%%%%%%%%%%%%%%%%%
\section{Introduction}

Stars habitually form in localised over-densities of cold gas, but predicting star formation (SF) remains challenging. Kennicutt (1998) proposed that the star formation rate (SFR) correlates principally with the total (i.e. \hi$+$\htw{}) gas surface mass density, with the gravitational collapse of sufficiently dense clouds in a turbulent ISM being the dominant SF mechanism (Mac-Low \& Glover 2012; Renaud et al. 2012). Alternatively, revisions of the SF Law in favour of a metallicity- and \htw-dependence have been advocated (Gao \& Solomon 2004; Bigiel et al. 2008; Gnedin \& Kravtsov 2011; Kuhlen et al. 2012). The apparent correlation of \htw{} and $\Sigma_{SFR}$ with mid-plane pressure in massive spirals (Wong \& Blitz 2002) implies that SF reflects the equilibrium between \htw-formation by dust grain catalysis and dissociation by the FUV (Far Ultra-Violet) background (Blitz \& Rosolowsky 2004). The significance of this mechanism, relative to turbulence-regulated cloud-collapse, is disputed, particularly in light of some resolved SF regions exhibiting very low \htw{} content (i.e. Boissier et al. 2008). 

These competing hypotheses both accommodate the inefficient SF found among metal-poor dwarves. On the one hand, a lack of or inefficiency of coolants in the ISM can restrict supersonic turbulence to higher density thresholds, thus mitigating the shocks that lead to SF (Renaud et al. 2012). On the other hand, weak \htw{} growth and protection from FUV in a metal-poor environment can push the \hi{}$-$\htw{} transition to higher surface densities (Kuhlen et al. 2012). 

Interstellar dust plays a seminal role in the latter mechanism, being responsible for shielding and \htw{} formation (Cazaux \& Tielens 2002). The properties of dust vary with environment (Cardelli et al. 1996), manifesting in a dependence on the local metallicity and quotient of Asymptotic Giant Branch (AGB) stars and Supernovae (SNe) Type 1a/II (Matsuura et al. 2009; Boyer et al. 2011). SNe II ejecta are believed to synthesise significant dust mass (Lisenfeld \& Ferrara 1998; Dwek 1998), and are the favoured culprits for high observed extinction in the early ($z>6$) universe (Dunne et al. 2003), although the efficiency with which ejected metals condense remains subject to debate given the simultaneous grain destruction in the reverse shock. Recent analytical models suggest SNe II are inefficient overall sources in dwarves, coupled with weak grain growth in the metal-poor ISM, leading to the high gas-to-dust ratios in these systems (Zhukovska 2014). Understanding the relative rates at which dust mass growth is contributed to by stellar sources and the ISM, as a function of metallicity, is key to elucidating conflicting observations of a {\it universal} dust-to-metal ratio (i.e. Zafar \& Wilson 2013; De Cia et al. 2013; Mattsson et al. 2014).

In this work, we examine the coupled \htw{} and dust lifecycles of the Magellanic Clouds over their recent ($z<0.3$) past. As uniquely gas-rich and blue galaxies within our Local Group, these neighbouring galaxies are a convenient probe of star formation more common at higher redshifts (Tollerud et al. 2011). The low differential rotation in the MCs also permits the direct association between star formation and local conditions such as supergiant shells (Leroy et al. 2009). Their proximity allows high resolution imaging of Giant Molecular Clouds (GMCs; Fukui et al. 2008), where almost 300 individual GMCs have been resolved from the NANTEN $^{\rm 12}$CO(1-0) survey of the LMC (Fukui \& Kawamura 2010).

%%%%%%%%%%%%%%%%%%%%%%%%%%%%%%%%%%%%%%%%%%%%%%%%%%%%%%%%%%%%%%%%%%%%%%%%%%%%%%%
\begin{table*}
\centering
%\begin{minipage}[t]{1.\columnwidth}
\caption{Summary of recent observations addressed in this work}
\begin{tabular}{@{}ccc@{}}
\hline
Stream & \hi{} mass/column density & Putman et al. (2003), Br\'{u}ns et al. (2005) \\ 
& Gas-Phase Metallicity & [S$_{\rm II}$/\hi]$\sim$ -1.1 (Fox et al. 2010, 2013) \\
& Depletion (dex) & $\sim$-0.6 (Fox et al. 2013) \\
& Gas-to-Dust ratio & 3.3$\sim$19 solar (Fox et al. 2013) \\ 
& Maximum stellar density & Br\'{u}ck \& Hawkins (1983) \\
\hline
Bridge/Wing & Gas-Phase metallicity & [Z/H]$=$-1.02 (Lehner et al. 2008) \\
& Stellar Metallicity & [Z/H]$=$-1.1 (Rolleston et al. 1999; Lee et al. 2005) \\
& \htw{} mass fraction & 0.002 (Mizuno et al. 2006) \\
& Dust-to-Gas Ratio & 1:1000 (Gordon et al. 2009) \\
\hline
SMC & SFH & Harris \& Zaritksy (2004) \\
& Current SFR (\md\yr{}) & 0.037-0.05 (Bolatto et al. 2011; Wilke et al. 2004) \\ 
& AMR & Harris \& Zaritsky (2004); Livanou et al. (2013) \\
& Metallicity Gradient & Cioni (2009) \\
& \htw{} mass (\md) & \ex{3.2}{7}{} (Leroy et al. 2007; 2009; 2011) \\
& Dust mass (\md) & 10$^{\rm 6}$ (Leroy et al. 2007; Bot et al. 2010) \\
\hline
LMC & SFH & Harris \& Zaritsky (2009); Weisz et al. (2013) \\
& Age-Metallicity Relation & Harris \& Zaritksy (2009); Rubele et al. (2012) \\
& Metallicity Gradient & Grocholski et al. (2006) \\
& \htw{} mass (\md) & \ex{4-10}{7}{} (Israel 1997; Fukui et al. 1999) \\
& Dust mass (\md) & \ex{1.6}{6}{} Matsuura et al. (2009) \\
\hline
\end{tabular}
%\end{minipage}
\end{table*}

The metallicity dependence of dust can also be addressed; at the mean metallicity of the SMC ($\sim$0.2 \zd{}; Bernard et al. 2008), the mass fraction of Polycyclic Aromatic Hydrocarbons (PAHs) decreases rapidly with metallicity (Smith et al. 2007). Sandstrom et al. (2012) attribute this drop to the inability of PAHs to survive the ambient UV medium, after having been formed in dense molecular clouds, with the implication that higher metallicity galaxies have softer UV fields {\it or} larger PAH grains. Moreover, the CO-to-\htw{} conversion factor is shown to vary substantially as a function of metallicity in the LMC and SMC (Leroy et al. 2011), with the SMC exhibiting a magnitude greater dissociation of CO relative to the \htw{} molecule. Dynamics or the stellar bar and gas inflow have been used to explain the observed age-metallicity relations (AMR) and radial profiles of metallicity (Cole et al. 2005; Carrera et al. 2008a, 2008b; Haschke et al. 2012a, 2012b); the correlation of the latter with the distribution of dust has also been detected in the LMC (Meixner et al. 2010). 

The diffuse \hi-dominated structures that envelope the Clouds (Putman et al. 2003; Nidever et al. 2008) provide a further test for present hypotheses of ISM evolution. The {\it Stream} is very metal poor (Fox et al. 2010, 13) with an oxygen abundance similar to that of the SMC when the Stream was presumed to have first formed ($\sim$2.5 Gyr ago; Nidever et al. 2010), and thus betraying its probable origin (Diaz \& Bekki 2012). No direct detection of dust or stars have been made within the $\sim200^{\circ}$  span of the Stream (Nidever et al. 2010), in spite of abundant \hi{} supply and an elemental depletion in the ISM of $\sim$0.6-0.7 dex comparable to the Galaxy (Fox et al. 2013). It is not known if the latter was established by {\it in situ} dust formation or is a signature of an earlier depletion level of the SMC. While dust could be preserved by the presumed lack of stellar feedback in the Stream, the interaction with the Galactic hot halo (Bland-Hawthorn et al. 2007) indicated by \ha{} would be hostile to grains. Conversely, the {\it Bridge}, a structure of similar tidal origin (Diaz \& Bekki 2012; DB12), offers strong evidence of extant molecular hydrogen (Muller, Staveley-Smith \& Zealey 2003; Mizuno et al. 2006), dust (Gordon et al. 2003), and young stars ($\sim$7 Myr old), formed {\it in situ} and not sourced from the SMC (Harris 2007). 

Recent data on the metals abundance of the Stream (Fox et al. 2013; Richter et al. 2013) however support the emerging hypothesis, previously suggested on the basis of kinematics, that the Stream is also sourced from the Large Magellanic Cloud (LMC). The exact mechanism for this is as yet undetermined: the presence of the Leading Arms (Putman et al. 2003) in conjunction with the trailing Stream is consistent with the classic tidal disruption of the SMC, but ram pressure interactions with the Galaxy can remove much of the LMC ISM in simulations (Mastropietro et al. 2005). Ionized High Velocity Clouds (HVCs) observed at the periphery of the LMC (Staveley-Smith et al. 2003) and matching it in kinematics and metallicity (Lehner et al. 2009) also imply strong stellar outflows (Olano 2004; Nidever et al. 2008). Elucidating the role of tidal interactions is however presently beset by large uncertainty in the orbital history of the MCs as derived from proper motions.

Recent developments in the numerical modelling of \htw{} formation, its conversion to new stars, and the accretion and destruction of multi-elemental dust, can help to address these issues. Previous sub-grid prescriptions for \htw-regulation have reproduced the empirical Kennicutt-Schmidt Law (Kennicutt 1998) for massive discs (Robertson \& Kravtsov 2008; Bekki 2013a) and dwarves (Kuhlen et al. 2012). A self-consistent treatment of dust accretion and destruction, introduced in Bekki (2013a), follows from semi-analytic models of the dust lifecycle (e.g. Dwek 1998; Lisenfeld \& Ferrara 1998; Hirashita et al. 2002). The complexity of the Magellanic System has thus far prohibited a detailed numerical study of the ISM, although a progression of increasingly refined dynamical simulations (i.e. Connors et al. 2006; Diaz \& Bekki 2012; Besla et al. 2012) currently support the tidal-dominated paradigm (Gardiner \& Noguchi 1996) for the formation of the Stream and Bridge. In a previous study, we explored how the diverse asymmetry of morphology and SF in Magellanic-type discs depends on tidal perturbations, halo structure and feedback (Yozin \& Bekki 2014). 

The principal aim of this paper, the first to numerically model the MCs' recent evolution with a multi-elemental representation of the ISM, is to combine the major phenomenological aspects of these earlier simulations and reproduce the salient multi-wavelength observations for star forming conditions of the MCs (Table 1). We also consider this a valuable test of our dust model. The paper is organized as follows: in the next section, we describe our parametrized structural models and \htw{}/dust model, derived largely from Yozin \& Bekki (2014) and Bekki (2013a; B13). The main results from several reference models and an accompanying parameter study are described in Section 3, and the applicability of these results to the Magellanic System and SF in a broader context are discussed in Section 4. Section 5 concludes this study.

%%%%%%%%%%%%%%%%%%%%%%%%%%%%%%%%%%%%%%%%%%%%%%%%%%%%%%%%%%%%%%%%%%%%%%%%%%%%%%%
\section{Numerical Method}

Our simulations commence with the backwards orbital integration (Murai \& Fujimoto 1980) of the MCs in proximity to the MW. The galaxies are represented with fixed mass point potentials, following trajectories consistent with their proper motions (DB12 and references therein). In a subsequent stage, the LMC or SMC potential is replaced with a live N-body model, and the simulation proceeds forward in time. For collisionless models and a conservative choice of proper motion (aligned with neither a first infall nor bound orbit scenario), DB12 exploited this method to a maximum lookback time of $\sim$3.4 Gyr, which they found sufficient to capture the tidal dynamics that reproduce the on-sky morphology. Conveniently, a period of elevated SF in the MCs (e.g. Weisz et al. 2013) is encompassed by this timescale. In this section, we provide an overview of these initial conditions and the sub-resolution physics recently adopted for these live models.

%%%%%%%%%%%%%%%%%%%%%%%%%%%%%%%%%%%%%%%%%%%%%%%%%%%%%%%%%%%%%%%%%%%%%%%%%%%%%%%
\subsection{Orbital Models}

For simplicity, the Galaxy is conveyed with a fixed mass potential. The dominant mass component is the extended halo, for which we adopt a NFW density distribution (Navarro et al. 1996), up to a viral radius $R_{cir}$ of 175 kpc:

\begin{equation}
\rho(r) = \frac{\rho_0}{(cr/R_{vir})(1+cr/R_{vir})^2}.
\end{equation}

In the present study, we use M$_{\rm vir}$$=$\ex{1.3}{12}\md{} and a virial radius R$_{\rm vir}$ = 175 kpc (Klypin et al. 2002). The Galactic disc uses the potential of Miyamoto \& Nagai (1975), which has the form:
\begin{equation}
\Phi_{disk} = -\frac{GM_d}{\sqrt{r^2 + (a + \sqrt{z^2 + b^2})^2}},
\end{equation}
where the total mass of the disk $M_d$ = 5.0$\times$10$^{\rm 10}$ \md{} (Binney \& Tremaine 2008), $a$ = 3.5 kpc and $b$ = 0.35 kpc. The Galactic bulge uses the spherical potential model of Hernquist (1990):
\begin{equation}
\Phi_{bulge} = -\frac{GM_b}{r+c_b}
\end{equation}
with mass $M_b$ is 0.5$\times$10$^{\rm 12}$ \md{} and $c_b$ = 0.7 kpc (Binney \& Tremaine 2008). For a total mass M$_{\rm MW}$ (within a limit radius of 300 kpc) of 1.73$\times$10$^{\rm 12}$ \md{}, the rotation profile is specified with regard to the Solar circular velocity of v$_{\rm cir}$ = 240 \km{} (i.e. Reid et al. 2009). 

The LMC and SMC potentials utilise the Plummer model:
\begin{equation}
\Phi_{mc} = -\frac{GM_{mc}}{\sqrt{r^2 + a_{mc}^2}},
\end{equation}
where M$_{\rm mc}$ is the total mass of the SMC/LMC (Section 2.2.1-2), $r$ is the distance from the centre of mass, and a$_{\rm mc}$ is the scale length. The MCs are subject to dynamical friction while within the massive halo of the Milky Way, which is implemented with the Chandrasekhar formula (Binney \& Tremaine 2008):
\begin{equation}
F_{d} = -\frac{4\pi G^2 M_{mc}^2 ln(\Lambda)\rho_{d}(r)}{\upsilon^2}[erf(X)-\frac{2X}{\sqrt{\pi}}exp(-X^2)]\frac{{\bf v}}{\upsilon},
\end{equation}
where M$_{\rm mc}$ is the total mass of the MC galaxy for whom the force is calculated, $\Lambda$ is 3.0 (Gardiner \& Noguchi 1996), $\upsilon$ is the velocity of the MC, and $X$ = $\upsilon/\sqrt(2\sigma)$, where $\sigma$ is an analytic approximation to  the isotropic velocity dispersion of the halo (Zentner \& Bullock 2003).

%%%%%%%%%%%%%%%%%%%%%%%%%%%%%%%%%%%%%%%%%%%%%%%%%%%%%%%%%%%%%%%%%%%%%%%%%%%%%%%
\begin{figure}
\includegraphics[width=1.\columnwidth]{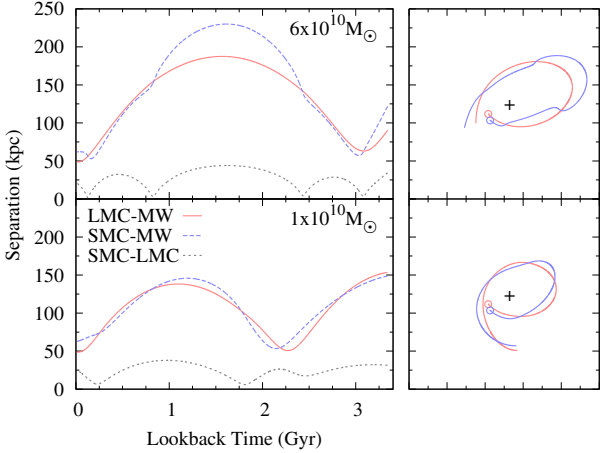} %sep.eps
\caption{(Left panels) Absolute distance between galaxies: LMC-MW (red line), SMC-MW (blue line) and LMC-SMC (black dashed line), as a function of lookback time. Top and bottom panels correspond to the high (L1; \ex{6}{10}{} \md) and low (L2; \ex{1}{10}{} \md) mass LMC models respectively (SMC and MW masses fixed in both cases). The orbits are constrained by observed positions and proper motions (Diaz \& Bekki 2012); (Right panels) Orbits of the LMC/SMC (red/blue) in the Y-Z plane, with present day locations indicated with open circles. The cross represents the MW centre of mass; plot side length is 500 kpc.} 
\label{fig_3}
\end{figure}

%%%%%%%%%%%%%%%%%%%%%%%%%%%%%%%%%%%%%%%%%%%%%%%%%%%%%%%%%%%%%%%%%%%%%%%%%%%%%%%
\subsection{N-Body Models}

We run N-body models with 10$^{\rm 6}$ particles, comprised initially of \ex{5}{5}{} DM, \ex{4}{5}{} old stars, and \ex{1}{5}{} gas particles treated with smooth particle hydrodynamics (SPH). The per-particle simulation time-step varies with density, lying within a range \ex{1.41}{4}{} to \ex{1.41}{6}{} yr. The gravitational softening length is provided individually for baryonic ($\eta_{\rm b}$) and dark matter ($\eta_{dm}$) particles; an average $(\eta_{\rm b}+\eta_{\rm dm})/2$ is utilised for baryon-dark matter interactions. All N-body simulations are implemented with an original code (e.g. B13), which executes with GPU clusters and the Pleiades computing architectures at the University of Western Australia, and gStar resources at Swinburne University of Technology

%%%%%%%%%%%%%%%%%%%%%%%%%%%%%%%%%%%%%%%%%%%%%%%%%%%%%%%%%%%%%%%%%%%%%%%%%%%%%%%
\subsubsection{SMC}

%%%%%%%%%%%%%%%%%%%%%%%%%%%%%%%%%%%%%%%%%%%%%%%%%%%%%%%%%%%%%%%%%%%%%%%%%%%%%%%
\begin{table}
\centering
\begin{minipage}[t]{1.\columnwidth}
\caption{Summary of initial conditions and parametrisations for the reference models}
\begin{tabular}{@{}cccccl@{}}
%\hline
& \multicolumn{2}{c}{LMC} & \multicolumn{1}{c}{SMC}  &\\
\hline
Parameter & L1 & L2 & S1 & Description \\
\hline
M$_{\rm t}$ (10$^{\rm 9}$\md) & 60 & 10 & 3 & Total Mass\\
M$_{\rm Halo}$/M$_{\rm Disc}$ & 9 & 2 & 1.5 & - \\
f$_{\rm g}$ & 0.3 & 0.3 & 0.8 & Gas mass fraction\\
Z$_{\rm o}$ & -0.6 & -0.6 & -1.0 & Central metallicity\\
$\alpha_{\rm m}$ (dex kpc$^{\rm -1}$) & -0.1 & -0.1 & -0.15 & Metallicity gradient\\
\hline
\tsn (yr) & \multicolumn{3}{c}{\ex{4}{5}}  & SN timescale \\
\rhoth (cm$^{\rm -3}$) & \multicolumn{3}{c}{10}  & SF threshold density\\
t$_{\rm acc,o}$ (yr) & \multicolumn{3}{c}{\ex{2.5}{7}}  & Accretion timescale\\
t$_{\rm dest}$/t$_{\rm acc}$ & \multicolumn{3}{c}{2}  & Destruction/accretion\\
\hline
\end{tabular}
\end{minipage}
\end{table}

Table 2 summarises the initial model parameters for our models. Our live N-body models for the formation of the SMC-sourced Stream are bound to the dynamical masses and orbital configuration of DB12. At \ex{3}{9}{} \md{}, the SMC mass conforms to earlier numerical studies on the origin of the Stream (Gardiner \& Noguchi 1996; Bekki \& Chiba 2005), and accommodates estimates from observed rotation curves (i.e. \ex{2}{9} \md{}, Westerlund 1997; \ex{2.4}{9}{} \md{}, Stanimirovic, Staveley-Smith \& Jones 2004). The observed stellar and \hi{} components have masses of $\sim$\ex{3.1}{8}{} \md{} (within r$<$3.5 kpc) and $\sim$\ex{5}{8}{} \md{} respectively (Staveley-Smith et al. 2003). Accordingly, DB12 utilise a halo-to-baryon mass \rh{} ratio of unity for their collisionless models, consistent with dwarves at a similar rotational velocity (Geha et al. 2006). In this work, we incorporate a dissipative component prone to destabilisation under substantial tidal interaction, and find M$_{\rm h}\ge\sim$1.5 is required to reproduce the on-sky morphology of the MCs. Besla et al. (2012) assume similar baryonic masses for their simulations commencing 7 Gyr ago, but use a halo-to-disc mass ratio on the order of 50 in accord with halo abundance matching for pre-infall SMC-analogues (Guo et al. 2010).

For this mass, we require a large initial gas fraction \fgas{} of 80 per cent, if asumming this initial quota includes material that later comprises the Stream/Bridge. The paucity of stars and the disturbed nature of tidal \hi{} makes distance estimates difficult, however, and provides the major uncertainty associated with estimates of the \hi{} mass (Stanimirovic et al. 2004). Brüns et al. (2005) suggest that the uncertainty regarding the distance of the Bridge, lying between the LMC at 50 and the SMC at 60 kpc, could result in a 20 per cent variation in assumed mass. Taking however the best estimates from the literature, the \hi{} contained within the SMC and its tidal structures is approximately 10$^{\rm 9}$ \md{}. 

These estimates may be lower limits, as they rely on an assumption of an optically thin 21cm line; Bernard et al. (2008) point out that if the observed emission arises from a colder component (T$\le$60 K), regions of high column density may conceal \hi{} masses $\sim$3.3 times larger. Taking a molecular hydrogen and Helium correction, and accounting for new stellar mass in this period of \ex{1-2}{8}{} \md, (plausibly inferred from the SFH; e.g. Harris \& Zaritsky 2004) the total gas fraction originally contained within the SMC could accumulate to more than \ex{1.5}{9}{} \md. By contrast, the present stellar mass is estimated to only \ex{3.1}{8}{} \md (Stanimirovic et al. 2004), implying a \fgas{} at 3 to 4 Gyr ago of $>$90 per cent; this far exceeds the average \fgas{} for low-luminosity dwarves (0.6; Geha et al. 2006). A large ionized gas component may need to be accounted for, as evidence for large corresponding fractions have been ascertained in regions as disparate as the Stream tip and Bridge (Lehner et al. 2008; Fox et al. 2010).

The initial disc size at 3.4 Gyr ago cannot be well constrained by the present size. Old stars have been found out to 6 kpc (N\'{o}el \& Gallart 2007) and De Propis et al. (2010) suggest this as the limiting radius; Nidever et al. (2011) find however Red Giants arranged almost axisymmetrically to 11 kpc. Moreover, Harris \& Zaritksy (2006) use kinematics of RGs to infer a primarily spheroidal structure with at least \ex{1.4}{9}{} \md{} within the central 1.6 kpc and up to \ex{5.1}{9}{} \md{} within 3 kpc. In addressing these observations, DB12 tested models of pressure-supported stellar spheroids embedded within the gas disc, for which the stripped kinematics and distribution compared favourably to those observed at the SMC periphery. Our focus on the ISM evolution justifies the simplification of the SMC model to a bulge-less disc with a NFW dark matter profile (Section 2.1), in which case the observed tidal morphology (Section 3.1) can be obtained with a disc size $R_{\rm d}=3.75$ kpc, with stellar scalelength ($R_{\rm 0}$) and scaleheight ($Z_{\rm 0}$) of 0.2$R_{\rm d}$ and 0.04$R_{\rm d}$ respectively.

Other aspects of the structural configuration are determined according the observed rotation curve and amplitude within the inner $\sim$4 kpc (Stanimirovic et al. 2004). The radial ($R$) and vertical ($Z$) density profiles of the stellar disc are assumed to be proportional to $\exp (-R/R_{0}) $ and to ${\rm sec}^2 (Z/Z_{0})$. The radial and azimuthal velocity dispersions are assigned with respect to epicyclic theory, with a Toomre parameter Q$=$1.5 providing an initially stable disc but susceptible to bar formation (Toomre 1963). The initial orientations of the disc ($\theta$=45$^{\circ}$ and $\phi$=230$^{\circ}$; Gardiner \& Noguchi (1996) obtain a present day viewing angle consistent with the SMC (Haschke, Grebel \& Duffau 2012a). Likewise for the LMC, we find $\theta$=-100$^{\circ}$ and $\phi$=260$^{\circ}$ are non-unique but satisfactory initial orientations, leading to the present viewing angle ($\sim$35$^{\rm \circ}$; van der Marel et al. 2002).

The initial distribution of metals is prescribed in a linearly varying fashion. At in-plane radius $R$ (kpc), the metallicity [m/H] (in dex) is defined in terms of a central metallicity [m/H]$_{\rm R=0}$ (dex) and gradient $\alpha_{\rm m}$ (dex kpc$^{\rm -1}$):
\begin{equation}
[m/H](R)=[m/H]_{R=0} + \alpha_m R,
\end{equation}

The Age-Metallicity Relation (AMR) of the SMC for stellar clusters (Da Costa \& Hatzidimitriou 1998; Piatti et al. 2001; Carrera et al. 2008a) suggests a mean enrichment of the ISM by $\sim$0.4 dex in the past 3-4 Gyr. This is commensurate with the metallicity of the Stream (0.1 \zd; Fox et al. 2013) at its presumed departure from the SMC disc several Gyr ago. The present metallicity gradient appears to be shallow across disparate stellar populations (Da Costa \& Hatzidimitriou 1998; Parisi et al. 2009; Cioni 2009), attributable to dilution encouraged by a strong stellar bar, or the regular infall/accretion of metal-poor gas triggered by tidal influence. Alternatively, a {\it small} primordial SMC, ostensibly without significant SF in the period bridging its initial formation ($>$12 Gyr ago) to the non-quiescent phase ($<$4 Gyr; Harris \& Zaritksy 2004), would be unlikely to retain a gradient in the presence of even secular mixing (Zaritsky et al. 1994). For this study, we explore $\alpha_m$ for our LMC and SMC models in the range -0.15 to -0.05 (dex kpc$^{\rm -1}$), in accordance with late-type gas discs at similar rotation velocity and mean metallicity (Zaritksy et al. 1994).

%%%%%%%%%%%%%%%%%%%%%%%%%%%%%%%%%%%%%%%%%%%%%%%%%%%%%%%%%%%%%%%%%%%%%%%%%%%%%%%
\subsubsection{LMC}

Van der Marel et al. (2002) finds a stellar mass for the LMC at \ex{2.7}{9}{} \md, using the mass-to-light ratio (with significant uncertainty) $M/L_V = 0.9\pm0.2$ (Bell \& de Jong 2001) together with the halo mass of \ex{8.7}{9}{} \md (within 8.9 kpc) found from Carbon Star kinematics. This motivated the orbital dynamics upon which DB12, and thus our SMC model, is based. Recent third-epoch data have however revised the dynamical mass within 8.5 kpc to 17$\pm$7$\times$10$^{9}$ \md{} (van der Marel \& Kallivayalil 2014). This higher mass lies closer in agreement with recent halo occupation models for pre-infall MC-mass analogues (Munshi et al. 2013), which imply a halo mass far larger, in some cases a magnitude, than those measured for the LMC and other Magellanic Spirals (i.e. Bush \& Wilcots 2004). 

To account for the uncertainty regarding the orbits (and corresponding halo stripping of the MCs within the Local Group since the commencement of our simulations 3.4 Gyr ago), we develop LMC models with dynamical masses of \ex{1}{10}{} and \ex{6}{10}{} \md, for which the baryonic mass (and halo radius) is determined from the reproduction of the rotation curve (van der Marel et al. 2002; Staveley-Smith et al. 2003), leading to \ex{3}{9}{} \md (1.5 R$_{\rm disc}$) and \ex{6}{9}\md (4 R$_{\rm disc}$) respectively. The present neutral component mass lies in the range of \ex{4-5}{8}{} \md (Kim et al. 1998; Staveley-Smith et al. 2003; Br\'{u}ns et al. 2005), assuming that the mass associated with the LMC and SMC can be distinguished from extraneous features by kinematics and column densities. By accounting for recent stellar growth (Harris \& Zaritksy 2009), minimal mass transfer to the less massive SMC, and \htw{} (10 per cent; Fukui et al. 2008) and He corrections, a \fgas{} of 30 per cent is reasonable at 3.4 Gyr ago. With a (truncated) stellar disc size of 7.5 kpc (i.e. Bekki \& Chiba 2005) and a gas disc 1.5 times more extended, our LMC model otherwise adopts the same mass distribution and kinematic model as the SMC (Section 2.2.1).

Figure 1 compares the orbits of the reference SMC (S1) and LMC (L1, L2) models, with respect to the Galaxy. The orbits of the low mass LMC (L2) and SMC are more consistent with a perigalacticon 1.8 to 2.5 Gyr ago, at which epoch the Stream is thought to have formed under tidal influence from the Galaxy. Besides the difficulty in obtaining high-precision transverse motions for the MCs, Zhang et al. (2012) show these orbits are highly sensitive to the solar circular velocity and halo structure of the Galaxy. A higher LMC mass also finds the SMC more tightly bound to it, leading to multiple close encounters. The similar SFHs of the MCs shed some light on these orbital encounters (Harris \& Zaritsky 2004; 2009), but the increasing uncertainty with lookback time presently precludes the precise determination of their occurrence and strength. Accompanying evidence of cluster formation at 2 Gyr ago and the formation of the Bridge $<$500 Myr ago support the occurrence of major interactions at these epochs (Piatti 2011), with which both LMC mass models approximately agree.

%%%%%%%%%%%%%%%%%%%%%%%%%%%%%%%%%%%%%%%%%%%%%%%%%%%%%%%%%%%%%%%%%%%%%%%%%%%%%%%
\subsubsection{Chemical enrichment and cooling}

Our model incorporates the evolution of 11 elements (H, He, C, N, O, Fe, Mg, Ca, Si, S and Ba). Chemical enrichment of the ISM relies on the expulsion of material from SNe 1a, SNe II and AGB stars; in each case, our code identifies the nearest gas particles as mass recipients. Our model implements the non-instantaneous recycling of chemical components in recognition of the time delay between massive star formation and the subsequent supernova event or AGB phase. Therefore the mass of each chemical element ejected from each stellar particle is time-dependent, and necessitates an approximation for the main-sequence turn-off mass from Renzini \& Buzzoni (1986). 

Bekki \& Tsujimoto (2012) finds the {\it prompt} SN Ia model to be necessary in explaining the chemical evolution of the LMC. This model consists of a time delay ($t_{Ia}$) between the SN Ia event and its metal ejection to the ISM, and is motivated by observational results (e.g., Manucci et al. 2006). We adopt a delay time distribution (DTD), for 0.1 Gyr $\leq t_{Ia} \leq $ 10 Gyr, which is consistent with the SN 1a rate in extra-Galactic galaxies (Totani et al. 2008). This rate is determined by the number of SN Ia per unit mass (which is controlled by the adopted IMF and the binary fraction for intermediate-mass stars for the adopted power-law slope of -1).

Radiative cooling, with a specified ratio of specific heat for the ISM $\eta = 5/3$, follows the rate curves of Rosen \& Bregman (1993) in the range $10^2 < T(K) < 10^4$, and Mappings III (Sutherland \& Dopita 1993) for $T > 10^4 K$. This cooling prescription in the critical low temperature regime is simplistic compared to recent studies (Maio et al. 2007) that indicate that the various Hydrogen-derived molecules (principally H, H$_{\rm 2}$, He and HD) are efficient coolants. 

%%%%%%%%%%%%%%%%%%%%%%%%%%%%%%%%%%%%%%%%%%%%%%%%%%%%%%%%%%%%%%%%%%%%%%%%%%%%%%%
\subsubsection{Dust Model}

Our dust model incorporates all gas phase elements heavier than He, with their formation restricted to SNe and AGB stars. For simplicity, we adopt an initially solar composition for metal and dust species, although the total metallicity is less than solar. This is justified in light of our primary focus in this preliminary study on the overall dust mass budget. To minimise the number of initial model parameters with high uncertainty, only {\it new} AGB stars and SN (those which form during the simulation) eject dust. While we acknowledge that this modelling may lead to an underestimation in dust mass, this is compatible with the tidal epoch of the MCs that we encompass in our simulations. The last 3.4 Gyr has seen elevated star formation in both galaxies (i.e. Weisz et al. 2013), in which case we can assume that the dust contribution by older stars is small compared to recent SNe/AGB output.

The total mass of stellar ejecta is estimated by using stellar yield tables by Tsujimoto et al. (1995) and van den Hoek \& Groenewegen (1997). The mass of metals locked in up in the dust phase, which varies with the element in question, is the same for SNe types 1a/II ($>$8\md), and varies for AGB (0.1\md$<$M$<$8\md) depending on the C/O fraction. A prominent but highly uncertain parameter in this model are the condensation efficiencies (mass fraction of ejected metals converted to dust grains) for each element, which we take from Dwek (1998) and were successfully adopted in B13 to reproduce massive spiral evolution. For SNe, the efficiency of (Mg, Si, Ca and Fe) and C are 0.8 and 0.5 respectively. For C-rich stars (C/O$>$1), the efficiency is unity for Carbon, and zero otherwise. bf We note that these efficiencies are high, and most applicable to evolved metal-rich systems. We discuss the implications in Section 4.1. Sulphur is treated as a special case with efficiency null, following Fox et al. (2013), who take metallicity measurements of the Stream ISM with [S$_{\rm I}$/\hi{}] as a reliable indicator due to its apparent zero depletion to the dust phase. 

%%%%%%%%%%%%%%%%%%%%%%%%%%%%%%%%%%%%%%%%%%%%%%%%%%%%%%%%%%%%%%%%%%%%%%%%%%%%%%%
\begin{figure}
\includegraphics[width=1.\columnwidth]{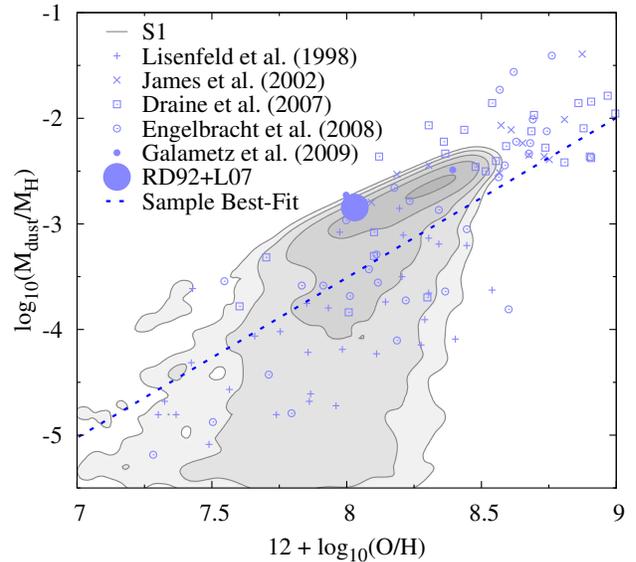}
\caption{Dust-to-gas ratio as a function of oxygen abundance. The contours and underlying logarithmic grayscale correspond to gas-phase particles of SMC model S1 lying within 3.5 kpc of the galaxy centre. The observational data provided for comparison include datasets for dwarf irregulars (Lisenfeld \& Ferrara 1998) and three low-metallicity galaxies of the more recent Galametz et al. (2009) sample. The blue dashed is a linear regression best fit to the aforementioned samples (Galametz et al. 2011). The blue circle indicates the intersection of the mean oxygen abundance for \hii{} regions and mean D for the SMC (Russell \& Dopita 1992; Leroy et al. 2007).} 
\end{figure}

Dust grain growth in the ISM is nominally dependent on an accretion timescale \tacc{}, which varies between gas particles, and is defined separately between elements. By adopting the one-zone approximation by Dwek (1998), which we simplify with the assumption of constant grain size and mass density, dust growth by accretion depends on the local gas density and thermal speed:
\begin{equation}
\tau_{acc}=\frac{\rho_0}{\rho}{\sqrt{\frac{T_0}{T}}}\tau_{acc,0},
\end{equation}
where $T$ and $\rho$ are the temperature and density of the respective gas particle from its SPH framework; $T_{\rm 0}$ and $\rho_{\rm 0}$ are typical Galactic values, equating to 20 K and 10 \cmc{} respectively. 

A reference local timescale corresponding to these conditions \tacco{} has no precedent in previous works. Dwek (1998) points out that it should be determined by recycling processes resident within low-mass molecular clouds which typically exhibit lifetimes of several 10$^{\rm 7}$ yr. In this work, we adopt a \tacco{} of \ex{2.5}{7}{} yr, which reproduces various observable properties of the MC ISM (Section 3).

Dust destruction occurs by sputtering (in which high energy atoms/ions, e.g. from SNe, collide with grains), background UV radiation, astration in stellar formation and outflows. Dwek (1998) noted that a spatial correlation in metallicity and the dust-to-gas ratio through the Galactic ISM implied a constant relationship between \tacc{} and the destruction timescale \tdest{}, irrespective of other environmental conditions. For the suggested \tdest$=2$\tacc{}, B13 was able to numerically reproduce the dust properties for secularly evolving massive galaxies, and thus we adopt this value, although this is clearly a major assumption for actively evolving metal-poor galaxies such as the MCs. Destruction occurs only in dust-contaminated gas particles that lie within a (SPH) smoothing length of SNe. For the astration of gas to new stellar particles, the gas-phase dust is effectively destroyed, contributing instead to the metal content of the new star.

We implement an initial dust-to-metal ratio (\rdm) of 0.4 (Dwek 1998). This value is consistent with the Galactic ISM, and other evolved systems at around solar metallicity in which condensation is efficient. In Section 4.1, we discuss the finding that the dust mass in our SMC and LMC simulations at zero lookback time is not sensitive to the adoption of an lower initial \rdm{} (i.e. 0.1, a value which is ostensibly more consistent with low metallicity systems; De Cia et al. 2013).

%%%%%%%%%%%%%%%%%%%%%%%%%%%%%%%%%%%%%%%%%%%%%%%%%%%%%%%%%%%%%%%%%%%%%%%%%%%%%%%
\subsubsection{Comparison with the observed D-\ao{} relation}

Figure 2 compares the dust-to-gas ratio (D) and the oxygen abundance (\ao{}) for our fiducial SMC model with the linear relation implied from observational samples. The simulated model data correspond to gas particles within 3.5 kpc of the SMC centre, and thus exclude most particles under significant tidal influence. The model data is broadly consistent with the observed global values for the SMC (Russell \& Dopita 1992; Leroy et al. 2007) as we similarly find for LMC model L1 and its observed counterpart. The slight excess of D and metallicity in our model compared to their observed counterparts is partly attributable to avoiding line-of-sight uncertainties by using only the dust and gas associated with each gas particle, but also assumptions regarding the initial \rdm{} and stellar dust production rate (Section 2.2.4), which we discuss in Section 4.1. Encouragingly, both models exhibit an deviation from linearity towards lower metallicity (similar to the observed transition from the linear regime at A$_{\rm O} = 8.1$ to 8.4; Smith et al. 2007). This non-linearity can be explained in terms of a smaller dust abundance, with reduced cloud opacity, enables more efficient destruction and thus less dust {\it etc}.

%%%%%%%%%%%%%%%%%%%%%%%%%%%%%%%%%%%%%%%%%%%%%%%%%%%%%%%%%%%%%%%%%%%%%%%%%%%%%%%
\subsubsection{H$_{\rm 2}$ formation and dissociation}

Formation and destruction of \htw{} relies on the balance of dissociating FUV radiation and its formation as a function of local dust, metallicity and kinematics. Our explicit modelling of dust provides a key distinction from Pelupessy, Papadopoulos, \& van der Werf (2006) upon which our \htw{} code is based. We are still limited however by mass resolution, such that GMCs (with typical mass ranging from 10$^{\rm 4}$ to 10$^{\rm 6}$ \md{}) and the variety of conditions within are not fully resolvable.

Grain catalysis (Cazaux \& Tielens 2002) is the primary driver of \htw{} formation, which we model self-consistently, albeit with the assumption of a constant formation efficiency per surface area for carbonaceous and silicate grains (Draine 2009). Using neutral and molecular hydrogen number densities estimated from our SPH framework, the rate of molecular formation can be approximated with a linear dependence (Jura 1975) on the dust-to-gas ratio of a gas particle (normalised by the Galactic value at 0.0064; Zubko et al. 2004). The relative influence of the ambient radiative field is approximated with an equilibrium model of Goldshmidt \& Sternberg (1995), which uses an efficient expression for \htw{} self-shielding (Pelupessy et al. 2006) and the radiation field. The intensity scaling factor for each gas particle is estimated from the cumulative flux of stars within 1000 \AA{} using the SEDs of Bruzual \& Charlot (2003) consistent with a Salpeter initial mass function (IMF).

%%%%%%%%%%%%%%%%%%%%%%%%%%%%%%%%%%%%%%%%%%%%%%%%%%%%%%%%%%%%%%%%%%%%%%%%%%%%%%%
\subsubsection{Star Formation and Feedback}

Stars form in dense cores of GMCs; the numerical application of this method is based on an earlier \hi-dependent code adopted originally by Katz (1992), where a gas particle is converted into a new star if (i) the local dynamical timescale is shorter than the sound crossing time scale (Jeans instability), (ii) the local velocity field is in gravitational collapse (i.e. div {\bf v}$<$0), and (iii) the local density exceeds a threshold \rhoth{}. The latter condition, here estimated using the molecular fraction of each gas particle \mfh{}, implicitly includes other physically-motivated SF criteria such as cooling rates and temperature. We also constrain the conversion of gas particles meeting the above criteria with a probability parameter, P$_{\rm SF}$, introduced in Bekki (2013b), to account for GMC masses lying below our model resolution:
\[
P_{SF}=1-exp(-\Delta t \rho^{\alpha_{sf}-1}),
\]
where $\Delta t$ is the timestep width for a given gas particle, $\rho$ is the density and $\alpha_{SF}$ is 1.5 (Kennicutt 1998). If P$_{\rm SF}$ exceeds a randomly generated number $R$ with 0$\le$R$\le$1, the conversion proceeds, based on the assumption that the timescale for star formation is shorter for higher density gas particles. 

Each stellar particle is formed with a fixed IMF and initial mass. The stellar mass decreases with time due to mass loss by SNe Ia, SNe II and AGB stars. The adopted IMF is prescribed with a power law (slope $\alpha$), and normalised to within the range 0.1 M$_{\odot}$ and 100 M$_{\odot}$. We adopt $\alpha$ = 2.35 for all models in the present study, corresponding to the Salpeter IMF, which has been shown to be valid for high mass (and therefore primary dust ejecting) stars (Sirianni et al. 2000). A fixed IMF is nonetheless a conservative choice for our dust- and \htw-centred work, given the presently indefinite relation between properties of cloud collapse and initial mass function (IMF). The bar dynamics and frequent infall-triggering events characteristic of the MCs will have likely led to a variety of star forming environments, with implications for the relative proportion of post-main sequence dust-forming populations.

%%%%%%%%%%%%%%%%%%%%%%%%%%%%%%%%%%%%%%%%%%%%%%%%%%%%%%%%%%%%%%%%%%%%%%%%%%%%%%%
\begin{figure}
\includegraphics[width=1.\columnwidth]{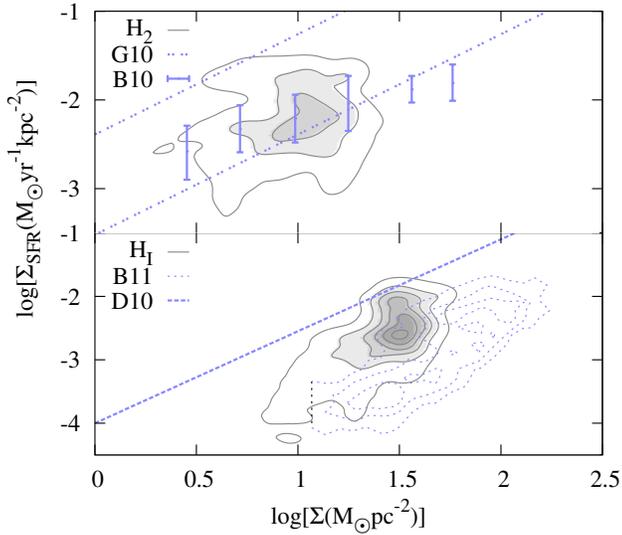}
\caption{(Bottom panel) The star formation rate surface density as a function of total (\hi$+$\htw{}) gas surface density. Grey contours, with linear scale, show the distribution from particles in the SMC model S1 at zero lookback time. A downsampling to 200 pc spatial resolution permits comparison with the  observed relation (dotted blue contours; Bolatto et al. 2011). The blue dashed line conveys the canonical ($\sim $1.4 slope) relation for spirals (Daddi et al. 2010); (Top panel) The SFR surface density with molecular hydrogen surface density; the grey contours correspond to model S1, blue bars from the observed SMC (Bolatto et al. 2011) and black dashed lines from best-fit relation for typical SF galaxies up to $z=3.5$ (Genzel et al. 2010).}
\end{figure}

Feedback is implemented by a SN model with the canonical $10^{51}$ ergs shared between thermal (UV emission) and mechanical kinematic components; the balance is controlled with the $f_{\rm k}$ parameter. A choice of $f_{\rm k}=0.1$ (10 per cent mechanical output, 90 per cent thermal) is consistent with observations of SNe and simulations of Thornton et al. (1998). The resulting shock extends to a radius of 0.175 kpc, with the energy distributed between neighbouring SPH particles. The timescale for the adiabatic phase of expansion (\ex{4}{5} yr) is established from fine-tuning of the models. In Magellanic-type galaxies at similar resolution, this range of $f_{kin}$ and \tsn{} mitigates spurious clumpiness (Yozin \& Bekki 2014). It was not sufficient however, to promote substantial SNe blowout, which Nidever et al. (2008) propose for the creation of the Leading Arm and one of two kinematically distinct filaments in the Stream.

%%%%%%%%%%%%%%%%%%%%%%%%%%%%%%%%%%%%%%%%%%%%%%%%%%%%%%%%%%%%%%%%%%%%%%%%%%%%%%%
\subsubsection{Comparison with observed star formation law}

Our results are based on simulations with \rhoth$ = $10 cm$^{\rm -3}$. This choice follows from physically-motivated thresholds within dense GMCs (i.e. transition to supersonic turbulence in the SMC ISM at 100 cm$^{\rm -3}$; Renaud et al. 2012), but also accounts for the mass resolution ($\sim$10$^{\rm 3}$ \md) which cannot resolve localised high density regions. This value is also constrained by the scale and distribution of \htw{}, where larger thresholds leave too much molecular hydrogen in the ISM, and too low a threshold can promote the {\it smearing out} of SF (Hopkins et al. 2013). We show in Section 3 that our choice of dust model and \rhoth{} leads to an appropriate mass of \htw{} and clustering of clumps, and confirm here that the simulated star formation law (Figure 3), for total gas- and \htw-surface density, compares acceptably with observed data from Bolatto et al. (2011), for whom $\Sigma_{\rm SFR}$ is derived indirectly from \ha{} and FIR emission, and assumes minimal obscuration. 

The \htw-dependence is reproduced particularly well, although we note our model makes no distinction between diffuse, cold, or CO-dark \htw{} components. We therefore predict a \mfh{} generally higher (at 10 to 15 per cent) than that directly detected in the SMC (10 per cent; Leroy et al. 2007). This can be reconciled with the uncertainty in X$_{\rm CO}$ of several factors at the SMC metallicity (Leroy et al. 2011), but we cannot ignore a systematic overprediction of $\Sigma_{\rm SFR}$, pushing our relation towards the observed trend among massive spirals and SF galaxies (Daddi et al. 2010; Genzel et al. 2010). We suspect this is a symptom of using a \htw{} and dust model and parametrization based on the Galactic-type ISM, for which the corresponding values at low metallicity are not well established.

%%%%%%%%%%%%%%%%%%%%%%%%%%%%%%%%%%%%%%%%%%%%%%%%%%%%%%%%%%%%%%%%%%%%%%%%%%%%%%%
\section{Results}

In this section we describe reference simulations for the LMC (L1, L2) and SMC (S1), with additional simulations illustrating parameter dependencies.

%%%%%%%%%%%%%%%%%%%%%%%%%%%%%%%%%%%%%%%%%%%%%%%%%%%%%%%%%%%%%%%%%%%%%%%%%%%%%%%
\subsection{Magellanic Stream}

%%%%%%%%%%%%%%%%%%%%%%%%%%%%%%%%%%%%%%%%%%%%%%%%%%%%%%%%%%%%%%%%%%%%%%%%%%%%%%%
\begin{figure}
\includegraphics[width=1.\columnwidth]{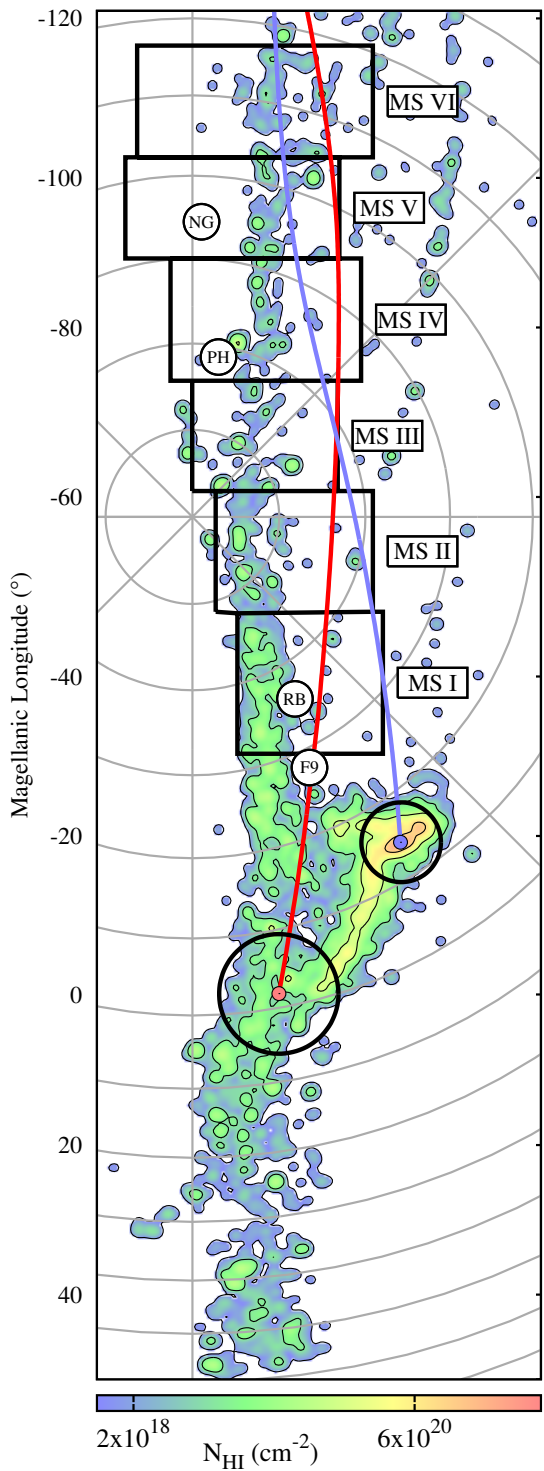}
\caption{\hi{} column density of the SMC model S1 at zero lookback time time in a ZEA projection. Straight lines give lines of galactic longitude (l) and curved lines give galactic latitude (b). The blue and red lines show the past orbits of S1 and the low mass LMC model L2 respectively. The limits of the logarithmic colour scale (\ex{1}{16}{} to \ex{1}{22}{} \cms) correspond to Fig. 8 of Nidever et al. (2010); the contours span the data of Putman et al. (2003), from their detection limit of $\sim$\ex{2}{18}{} \cms{} to \ex{6}{20}{} \cms{}. The MS regions I to VI (Mathewson et al. 1974) are shown in thick black boxes; the AGN sightlines for metallicity sampling of the Stream (Fox et al. 2013) are represented by white circles.}
\end{figure}

Simulation S1 was constrained to reproduce the salient features of the Stream/Bridge. Our model is the first to explicitly model \hi{}, permitting a direct comparison with observation. Figure 4 shows the \hi{} column density of S1 with a projection centred on the South Galactic Pole and with the $y$-axis representing the Magellanic longitude (Wakker 2001). The per-pixel density was derived from simulation data by projecting sightlines with width consistent with the effective resolution ($\sim$16$^{\circ}$) of the Parkes beam. The calculation also assumes the distance of the SMC, LMC and all other extraneous material lie at heliocentric distances of 60, 50 and 55 kpc respectively (Putman et al. 2003). 

In practice, all components show considerable line-of-sight depth; for example, radial velocities can distinguish two distinct filaments extending from the root of the Stream, with one obscuring the other in our projection (Section 3.1.6). The apparent mechanism behind this {\it bifurcation}, as interpreted by Diaz \& Bekki (2012), is the wrapping of a SMC tidal arm by the MW {\it after} having been originally stripped by the LMC/MW. In this scenario, the initial emergence of the tidal arm from the disc is dated to ($\sim$1.5 to 2 Gyr) ago, as in Gardiner \& Noguchi (1996). The Bridge, emerging from the recent (<300 Myr) LMC-SMC orbit pericentre, is clearly apparent as an \hi{} filament drawn towards the LMC, resulting in non-negligible accretion to the LMC disc. As in the collisionless models of DB12, the leading \hi{} tidal component stripped from the SMC lies in advance of the LMC orbit (red line) rather than the more broad distribution of the observed Leading Arms with a 60$^\circ$ deflection from the Stream (Putman et al. 1998). 

%%%%%%%%%%%%%%%%%%%%%%%%%%%%%%%%%%%%%%%%%%%%%%%%%%%%%%%%%%%%%%%%%%%%%%%%%%%%%%%
\begin{figure}
\includegraphics[width=1.\columnwidth]{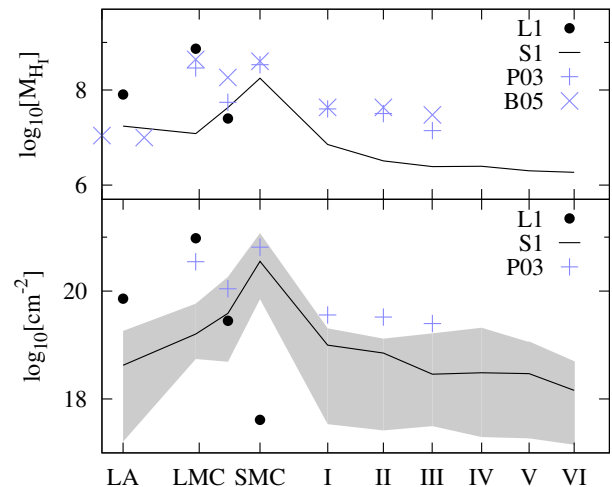}
\caption{The mass composition of the Magellanic system as a function of Magellanic longitude. (Bottom) The simulated \hi{} column density of S1 and L1 (shaded region represents 1$\sigma$ dispersion for S1) compared with the mean observed values (Putman et al. 2003), provided for the regions I to VI (Figure 4); (Top) the total \hi{} mass, compared with data from Putman et al. (2003) and Br\'{u}ns et al. (2005). The simulations compare well in the vicinty of the MCs, but the majority of the Stream conveys up to a magnitude deficit in mass.}
\end{figure}

This SMC-sourced Stream is generally coincident with its observed location in regions I to VI (Mathewson et al. 1974) and sightlines (Fairall 9, RBS 144) near the root (Fox et al. 2013). The offset of the Stream from the past SMC orbit (blue line) where it joins the LMC-SMC \hi{} envelope is due to strong tidal action by the LMC. This can also explain the poor spatial consistency towards the Leading Arm and the overly massive and dispersed Stream tip, although ram pressure is a further possibility not accounted for here. Mastropietro et al. (2005) reproduce the location of the Stream, and a density gradient with Magellanic Longitude (Westerlund 1997), primarily through ram-pressure interaction with the MW. 

%%%%%%%%%%%%%%%%%%%%%%%%%%%%%%%%%%%%%%%%%%%%%%%%%%%%%%%%%%%%%%%%%%%%%%%%%%%%%%%
\subsubsection{\hi{} Mass}

We find a minor gradient in \hi{} mass and mean column density across the Stream (Figure 5); this generally occurs in our models where the bulk of the Stream is not stripped too early by the LMC before being tidally disrupted by the MW following perigalacticon (2 to 2.5 Gyr ago). Our capacity to constrain the relative tidal forces between the LMC-SMC is limited however by a factor $\sim$5 deficit in the mean column density found along the Stream (an issue also faced by the more massive models of Besla et al. 2012), when compared with Putman et al. (2003) and Br\'{u}ns et al. (2005). The corresponding total \hi{} mass is distance-independent and similarly shows this deficit. At the fixed \msmc$=$\ex{3}{9}{} \md{} mass, increasing the initial gas quotient to placate this disparity only leads to excessive mass stripping leading to a parent SMC body no longer following the orbital dynamics of DB12 (leading to an erroneous on-sky distribution). Mitigating this process through further compaction of the initial gas/stellar disc leads to rapid gas depletion and enrichment not consistent with the AMR of SMC clusters (i.e. Harris \& Zaritsky 2004; Livanou et al. 2013), while further compaction of the halo prevents the tidally-induced bar instability from propagating.

%%%%%%%%%%%%%%%%%%%%%%%%%%%%%%%%%%%%%%%%%%%%%%%%%%%%%%%%%%%%%%%%%%%%%%%%%%%%%%%
\begin{figure}
\includegraphics[width=1.\columnwidth]{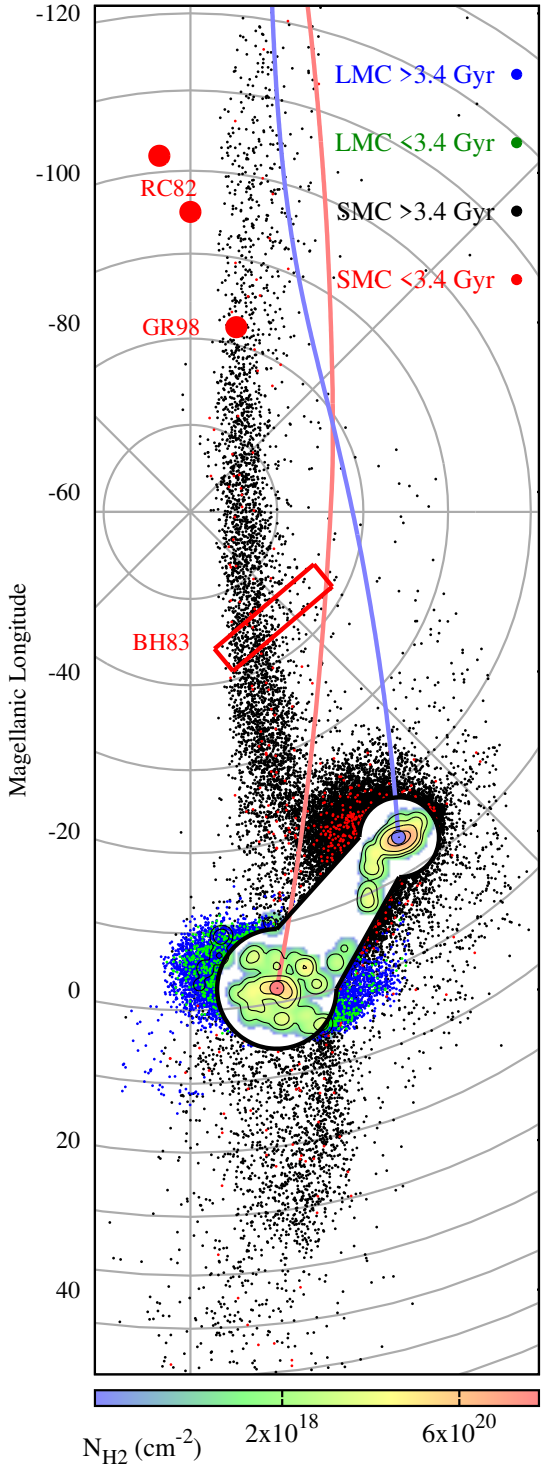}
\caption{A ZEA projection of the distribution of old (age $>$ 3.4 Gyr) and new (age $<$ 3.4 Gyr) stars originating from L1 (blue and green dots respectively) and S1 (black and red dots respectively). The stellar counterpart to the Stream is dominated by old SMC-sourced stars. The projected regions corresponding to the LMC, SMC and Bridge are masked out. The underlying colour scale shows the \htw{} column density from \ex{2}{16} to \ex{6}{20}{} \cms. Within this range, \htw{} does not appear in the Stream/Leading Arm. For comparison, the regions previously surveyed for stellar material i.e. BH83 (Br\'{u}ck \& Hawkins 1983); RC82 (Recillas-Cruz 1982) and GR98 (Guthathakurta \& Reitzel 1998) are indicated in red.}
\end{figure}

%%%%%%%%%%%%%%%%%%%%%%%%%%%%%%%%%%%%%%%%%%%%%%%%%%%%%%%%%%%%%%%%%%%%%%%%%%%%%%%
\begin{figure}
\includegraphics[width=1.\columnwidth]{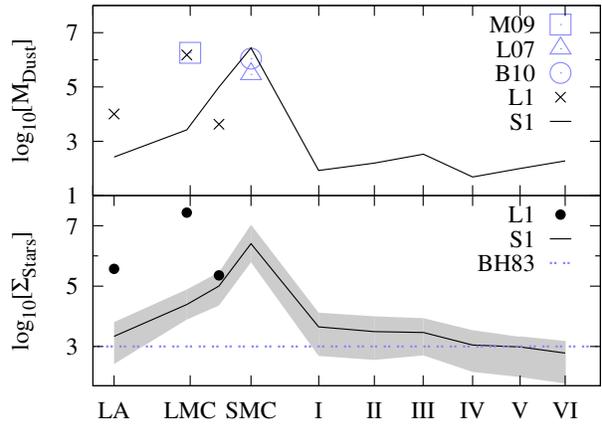}
\caption{Mass composition of the Magellanic system as a function of Magellanic longitude. (Bottom) The mean surface mass density of stellar particles, compared with the upper limit proposed by Br\'{u}ck \& Hawkins (1983); (Top) the total dust mass, compared with observationally inferred estimates for the LMC from Matsuura et al. (2009), and for the SMC from estimates of Leroy et al. (2007; blue triangle) and Bot et al. (2010; blue circle).}
\end{figure}

%%%%%%%%%%%%%%%%%%%%%%%%%%%%%%%%%%%%%%%%%%%%%%%%%%%%%%%%%%%%%%%%%%%%%%%%%%%%%%%
\subsubsection{Stellar content of Stream}

Material stripped from our reference LMC (L1; \mlmc$=$\ex{6}{10}{} \md) does not contribute significantly to the Stream, in contrast to the scenarios posited by Mastropietro et al. (2005) and Nidever et al. (2008). An alternative explanation for the deficit is that the total \hi{} mass of the Stream cannot be well constrained due to an absence of detected stars for a reference distance. There have been various unsuccessful attempts to find stars in the Stream (e.g. Philip 1973; Ostheimer et al. 1997). Figure 6 shows the distribution of old and new (age $<$ 3.4 Gyr) stellar particles, at a mass resolution of $\sim$10$^{\rm 3}$\md. The stellar counterpart to our modelled Stream, dominated by old stars, traces the same location as the \hi{} (Figure 4), and exhibits a faint surface mass density gradient (Figure 7). The unsuccessful survey of Guhathakurta \& Reitzel (1998) is coincident with this predicted stellar stream, while that of Recillas-Cruz (1982) is not. 

Brück \& Hawkins (1983) predict a maximum on-sky density of 1000 per square degree towards the base of the Stream. For model S1, mean stellar densities along the Magellanic Longitude lie between 10$^{\rm 3}$ and 10$^{\rm 4}$ \md{} per square degree. Our results are thus broadly consistent with the limited availability of observational data, and commensurate with the expectation of inefficient stripping to the Stream (Yoshizawa \& Noguchi 2003), where most stripped stars reside in a stellar halo instead (Weinberg 2000). Diaz \& Bekki (2012) find a similar mean density for models with a spheroidal stellar component of radius 5 kpc; a 2.5 kpc radii leads to a magnitude reduction in the projected stellar density.

%%%%%%%%%%%%%%%%%%%%%%%%%%%%%%%%%%%%%%%%%%%%%%%%%%%%%%%%%%%%%%%%%%%%%%%%%%%%%%%
\subsubsection{Dust and \htw{} masses}

The \htw{} content of our simulated SMC is consistent with recent CO-independent measurements of $\sim$\ex{3}{7}{} \md{} (Leroy et al. 2007; Bolatto et al. 2011). The underlying colour scale of Figure 6 shows \htw{} to be concentrated within \hi-rich regions of Figure 4, in particular the SMC Wing and southern branch of the Bridge. Given the lower limit on the logarithmic colour scale of $\sim$\ex{2}{16}{} \cms, there appears to be negligible \htw{} in the Stream and leading tidal features, compared with a mean $\sim$10 per cent in the SMC parent body (Israel 1997). 

The total dust content in S1 (Figure 7) lies towards the upper limit established from IR and sub-millimetre imaging, which estimate a mass of \ex{(0.29-1.1)}{6}{} \md{} (Leroy et al. 2007; Bot et al. 2010; Zhukovska \& Henning 2013). Large dust-to-gas ratios of clusters within the SMC tail (Gordon et al. 2009), formed from tidal action on the SMC, suggest that some dust can be stripped and persist in low metallicity environments for several Myr. Our model cannot specifically address the dust-to-gas ratio at the Fairall 9 sightline which Fox et al. (2013) suggests exceeds that of the SMC. Mathewson et al. (1979) similarly proposed the existence of obscuring material near peak \hi{} regions in the Stream, based on galaxy counts, but dust was not detected in the early IRAS IR maps of the Stream (Fong et al. 1987). These results rely on the assumption that the dust is well mixed with the gas, and is warm (>15 K) by means of friction with the MW halo and ambient radiation field, if not ionising stars within the Stream. 

%%%%%%%%%%%%%%%%%%%%%%%%%%%%%%%%%%%%%%%%%%%%%%%%%%%%%%%%%%%%%%%%%%%%%%%%%%%%%%%
\subsubsection{D and elemental depletion in the SMC-sourced Stream}

Figure 8 shows the metallicity and dust-to-gas ratio for the ISM of S1 lying along the line-of-sight as a function of Magellanic Longitude. Metal abundance is conveyed in Sulphur and Iron ratios (relative to Solar; Asplund et al. 2009), to provide comparison with Fox et al. (2013), who establish the metallicity of the Stream from several AGN sightlines (shown in Figure 4). They find minimal depletion of the gas-phase $[S_{\rm II}/H_{\rm II}]$ ratio in the presence of dust/ionization, and accordingly obtain a metallicity of $\sim$0.1 \zd{}, spanning from the root up to -100$^\circ$. A notable exception lies in the direction Fairall 9 (corresponding to one of two kinematically distinguished filaments at the root; Nidever et al. 2008) whose 5 times greater metal abundance cannot be accounted for simply by beam-size uncertainty over such small-scales. The authors conclude that the lower metallicity filament originates from the SMC based on the {\it mean} metallicity implied from the AMR at 1.5 Gyr ago. 

We show here the first numerical validation for this argument, based on an initial metallicity profile that also satisfies the AMR in the inner SMC disc (Section 3.2.2). The initial abundance gradient $\alpha_{\rm m}$$=$-0.15, over the 7.5 kpc radial span of the S1 gas disc, is imprinted on the Stream by means of the gradual decline in $[S/H]$ with distance from the SMC. Moreover, the dust and metal abundance of the leading tidal plume are similar to the Stream, as suggested by early chemical data (Lu et al. 1998; Gibson et al. 2000), and in support for the tidal origin from the SMC. The corresponding spread of the S1 ISM in [S/H] in the vicinity of Fairall 9 of around 0.1 dex deems it unlikely however to accommodate the relatively metal-rich filament coinciding with Fairall 9 (Fox et al. 2013).

%%%%%%%%%%%%%%%%%%%%%%%%%%%%%%%%%%%%%%%%%%%%%%%%%%%%%%%%%%%%%%%%%%%%%%%%%%%%%%%
\begin{figure}
\includegraphics[width=1.\columnwidth]{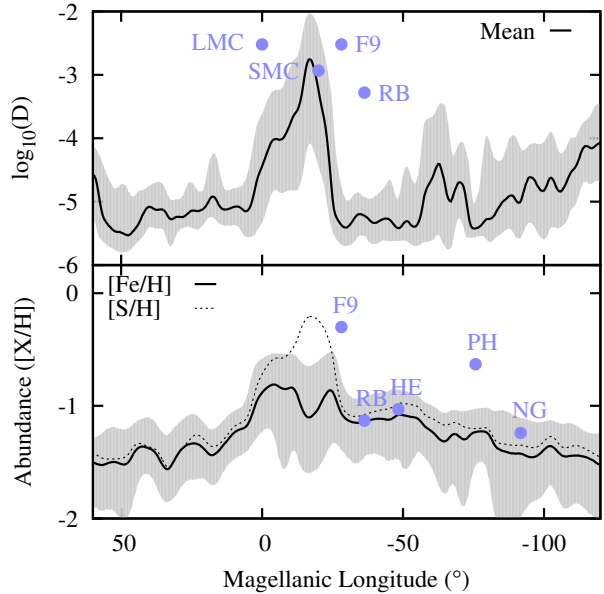}
\caption{Gas-phase metallicity and dust-to-gas ratio of the SMC-sourced Stream from model S1, as a function of Magellanic Longitude. (Bottom panel) The mass weighted mean [Fe/H] along {\it l} is shown with the thick black line (shaded region represents 1$\sigma$ dispersion); the equivalent mean for [S/H] is shown with the  dotted black line. The sightlines of Fox et al. (2013) used for metallicity analysis of the Stream are shown i.e. F9 (Fairall 9); RB (RBS 144); HE (HE0056-3622); PH (PHL 2525); NG (NGC 7714); (Top panel) Dust-to-gas ratio of S1 (black line, with 1$\sigma$ dispersion in grey), compared with blue dots representing RBS 144 (RB) and Fairall 9 (F9), the SMC (Leroy et al. 2007, Gordon et al. 2009) and LMC (Draine 2003).}
\end{figure}

The mean gas-phase depletion of the simulated stream is [Fe/H]$<$0.1 dex, inconsistent with the observed $\sim$0.6 dex. By comparison, the parent SMC disc, with ongoing star formation and the presence of \htw{}, is consistent with Welty et al. (2001) who find, in the SMC {\it wing}, dust depletions of Fe/Ni/Zn to vary from -0.3 dex to -1.7 dex in low and high column densities respectively.

Figure 9 shows the spatial distribution of D in the same projection as Figure 4; the SMC disc and tail are consistent with observation (i.e Gordon et al. 2009). At locations outside the SMC/Bridge, D is largely homogenous at $\sim$10$^{\rm -5}$, in accord with the initial D at the periphery of the gas disc, which in turn was set by the initial metallicity, metallicity gradient and the dust-to-metal ratio. The complex dynamics of the interaction leads to very localised regions with D up to 10$^{\rm -3}$. It is difficult however to reconcile these with a coherent dust-rich filament, as implied by Fox et al. 2013), which possesses an upper limit on D only a factor of two smaller than the SMC (Leroy et al. 2007); we discuss this result in Section 4.1.3. 

%%%%%%%%%%%%%%%%%%%%%%%%%%%%%%%%%%%%%%%%%%%%%%%%%%%%%%%%%%%%%%%%%%%%%%%%%%%%%%%
\begin{figure}
\includegraphics[width=1.\columnwidth]{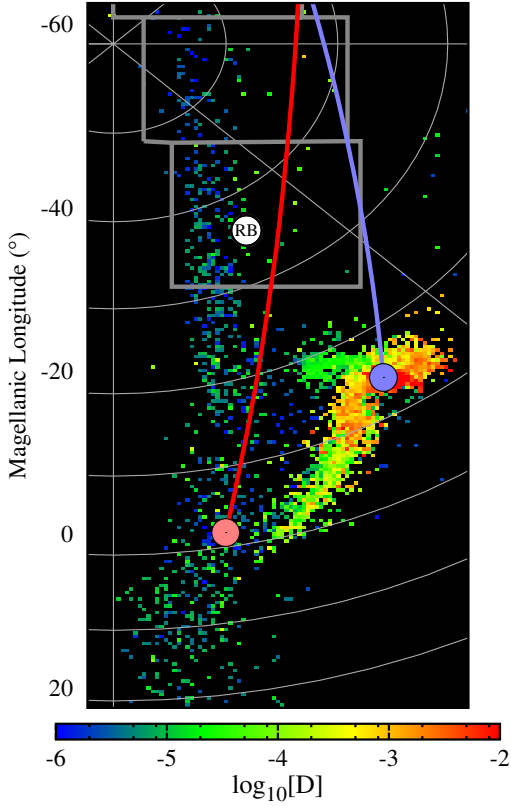}
\caption{Dust-to-gas ratio of SMC model S1, in the same projection as Figure 4. The colour scale is logarithmic, ranging from 10$^{\rm -6}$ to 10$^{\rm -2}$. RB represents the location of the metal-rich filament suggested by Fox et al. (2013). The red (blue) circle and line represent the LMC (SMC) and its past orbit respectively.}
\end{figure}

%%%%%%%%%%%%%%%%%%%%%%%%%%%%%%%%%%%%%%%%%%%%%%%%%%%%%%%%%%%%%%%%%%%%%%%%%%%%%%%
\subsubsection{Promoting depletion in the SMC-sourced Stream}

%%%%%%%%%%%%%%%%%%%%%%%%%%%%%%%%%%%%%%%%%%%%%%%%%%%%%%%%%%%%%%%%%%%%%%%%%%%%%%%
\begin{figure}
\includegraphics[width=1.\columnwidth]{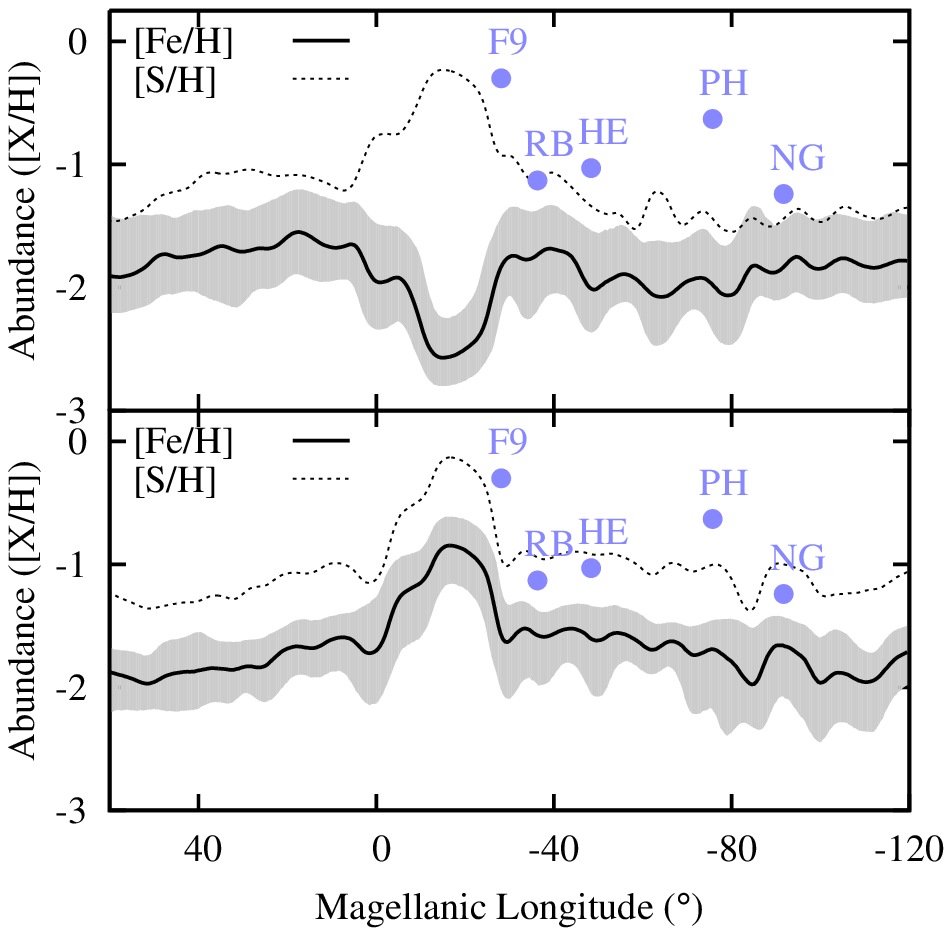}
\caption{Mean Sulphur (dotted black line) and Iron (thick black line, with shaded region representing 1$\sigma$ dispersion) abundance of the Magellanic System as a function of Magellanic Longitude, with sightline measurements from Fox et al. (2013) highlighted with blue circles. The bottom panel shows the results from the model with \tsn$=$\ex{1}{5}{} yr and \rhoth$=$50 \cmc). The top panel shows the same for an environment-independent dust lifecycle model, whose non-variable dust accretion timescale, established from Galactic models (0.25 Gyr; Dwek 1998; B13) naturally reproduces the observed $\sim$0.6 dex depletion.}
\end{figure}

Figure 10 shows the iron and sulphur abundances of the simulated SMC-sourced Stream, in non-reference simulations in which fine-tuning of model parameters attain the observed $\sim$0.6 dex depletion level in the Stream. The top panel corresponds to the fixed dust timescale case of B13, where dust growth/destruction are independent of local conditions. We use \tacco$=$0.25 Gyr and \tdesto$=$0.5 Gyr, derived from Dwek (1998) from the Galaxy, and successfully adopted by B13 to reproduce the main observed dust properties of massive spirals. This prescription leads to a constant depletion level in both parent disc and the most diffuse regions of the tails. As noted by Dwek (1998), this is a direct consequence of the proportionality in formation/destruction timescales. The corresponding dust-to-gas ratio ($\sim$0.001) in the vicinity of sightlines Fairall 9 and RBS 144 (Fox et al. 2013) matches their observed D with respect to a Galactic value of 0.01 (Knapp \& Kerr 1974).

In reference S1, we found iron depletion is concentrated towards the disc centre (Figure 8), where gas is directed and SF is most efficient (leading to a 1 dex depletion in the disc nucleus). An alternative (but non-unique) means of promoting tail depletion (bottom panel, Figure 10) follows from decreasing the SN expansion timescale four-fold (\tsn$=$\ex{1}{5}{} yr) and increasing the SF threshold parameter five-fold (\rhoth$=$50 \cmc). The combination of changes described improve the retention of \htw{}, and thus endorse depletion within the outer disc, prior to its stripping. The Stream depletion thus meets the observed level and remains preserved in the largely passive ISM of the Stream/leading arm. A major caveat to this model lies however in the excessive \htw{} mass, and overall enrichment inconsistent with the observed AMR (Section 3.2.2). 

%%%%%%%%%%%%%%%%%%%%%%%%%%%%%%%%%%%%%%%%%%%%%%%%%%%%%%%%%%%%%%%%%%%%%%%%%%%%%%%
\subsubsection{The hypothetical LMC-filament in the Stream}

%%%%%%%%%%%%%%%%%%%%%%%%%%%%%%%%%%%%%%%%%%%%%%%%%%%%%%%%%%%%%%%%%%%%%%%%%%%%%%%
\begin{figure}
\includegraphics[width=1.\columnwidth]{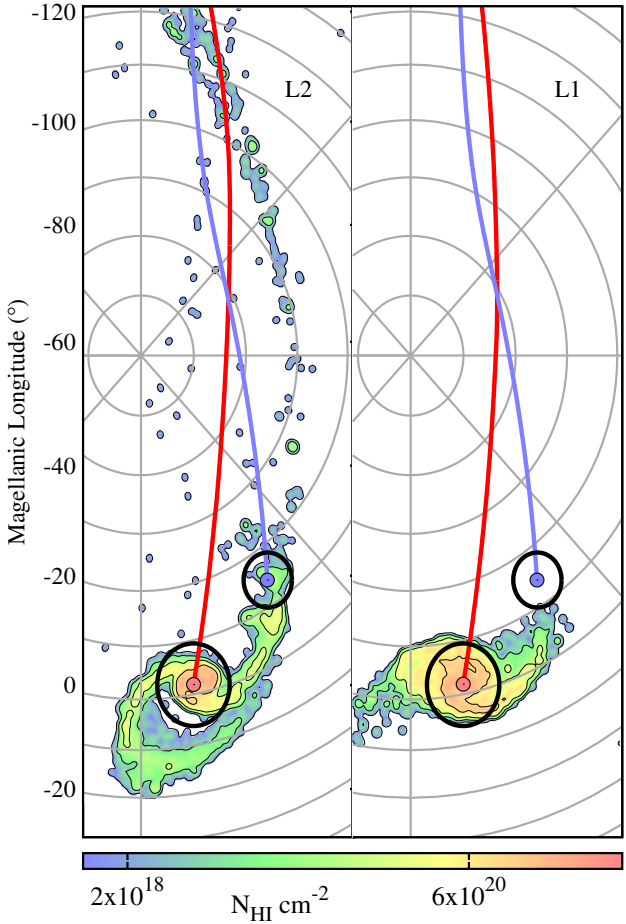}
\caption{Low mass LMC model L2 (left) and high mass L1 (right) projected in the same manner as in Fig. 4.} 
\end{figure}

Figure 11 compares the on-sky morphology of low and high mass LMC models (1 and \ex{6}{10}{} \md respectively) in the same projection as Figure 4. The stripping of the LMC in a large LMC-SMC mass ratio (20:1) is confined to dual tidal arms extending up to 10 kpc from the parent disc (right panel). The observed LMC in fact shows only a single tidal arm/filament. Ram pressure interactions have been shown to sufficiently truncate the simulated gas disc (Mastropietro et al. 2005); the implication is that the eastern arm was dissipated by these interactions while the western arm is shielded within the LMC-SMC interface region. In simulations without the SMC, we find similar tidal arms emerge as a result of MW stripping only, and thus the alignment of one arm in with the LMC-SMC Bridge may be serendipitous, and not necessarily related to stripping by the SMC. The left panel of Figure 11 shows that a low mass LMC (with LMC-SMC mass ratio 10:0.3) is sufficiently perturbed by the MW and SMC to leave a substantial \hi{} tail, stretching as far as the observed Stream (Nidever et al. 2010). The morphology of the LMC-sourced tail is clearly not coincident with the observed Stream, with its curvature continuous with that of the Bridge.

We can therefore apply the argument, as adopted for the SMC-sourced Stream, that the LMC-sourced tail would be imprinted with the metallicity of the LMC at this earlier epoch (i.e. Fox et al. 2013). The mean metallicity of L1 at 1.5 to 2 Gyr was $\sim-0.5$ (Section 3.2.2), but the LMC-sourced stream has [Fe/H]$\sim-1.0$. This is consistent with the least bound and most metal-poor periphery of the LMC gas disc, but is quantifiably similar to the SMC-sourced Stream thus making it difficult to distinguish on metallicity alone. The metal-rich filament at [Fe/H]$=0.3$, coincident with Fairall 9 (F9), is not reproduced either in this case or even in models with a very shallow initial metallicity gradient (i.e. $\alpha_{\rm m}$$=$-0.05 dex kpc$^{\rm -1}$). 

%%%%%%%%%%%%%%%%%%%%%%%%%%%%%%%%%%%%%%%%%%%%%%%%%%%%%%%%%%%%%%%%%%%%%%%%%%%%%%%
\begin{figure}
\includegraphics[width=1.\columnwidth]{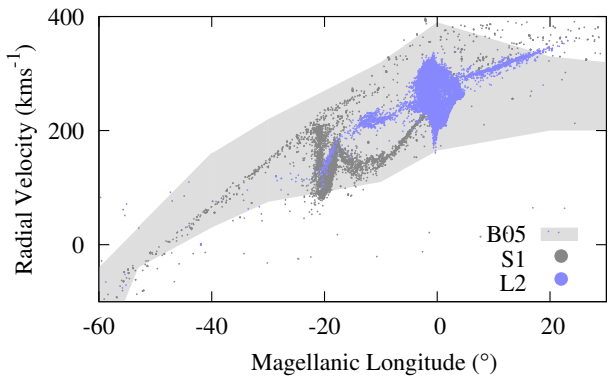}
\caption{The line-of-sight (radial) velocity (kms$^{\rm -1}$) of low mass LMC model L2 (blue points) and SMC model S1 (grey dots), as a function of Magellanic Longitude. The shaded region is the observational limits of Br\'{u}ns et al. (2005).} 
\end{figure}

The \hi{} mass (and column density) of this tail is generally a magnitude less than the SMC equivalent (Figure 5); the stellar tail counterpart traces a similar on-sky location, at a stellar density less than the SMC and thus not inconsistent with the upper limit estimated by Br\'{u}ck \& Hawkins (1983). Our models thus indicate that the expulsion of the LMC ISM, in the form of Supergiant Shell blowout as speculated in previous works (Olano 2004; Nidever et al. 2008), is not strictly necessary to explain the formation of an \hi{} stream with no stellar counterpart.

Figure 12 shows how the simulated radial velocity profile in the LMC, SMC and trailing tidal features is commensurate with the observed range (Br\'{u}ns et al. 2005). The principle inconsistency lies in the kinematics of the leading arms ($l_{\rm MC}>0^{\circ}$), although ram pressure would plausibly fragment these structures, which lie further into the MW potential well. Within the Bridge (-20$^{\circ} < l_{\rm MC} <$ 0$^{\circ}$) and interface region at the Stream root (-30$^{\circ} < l_{\rm MC} <$ -20$^{\circ}$), the LMC and SMC components exhibit distinct radial velocities. The presence of this {\it bifurcation} is noted in observations (Putman et al. 2003; Nidever et al. 2008), who find two spatially and kinematically distinct \hi{} filaments in the range -40$^{\circ} < l_{\rm MC} <$ -5$^{\circ}$. 

One of these filaments can be traced to the LMC, specifically, the south-east \hi{} over-density (SEHO) that houses both the prominent star forming region 30 Doradus and the Molecular Ridge (Ott et al. 2008). Nidever et al. (2008) propose that \hi{} gas was forced out of this SEHO to form a filament that provides the bulk of the northern branch of the Bridge, and the more kinematically energetic Stream filament. This hypothesis is supported by an apparent periodicity in the radial velocity, corresponding to the removal of gaseous material from the SEHO within the rotating LMC gas disc. They adopt this scenario to date the 200 degree long Stream to at least 2.5 Gyr old. This hypothesis is more plausible than the interpretation of the {\it helical} nature of the Stream (Putman et al. 2003) as a consequence of a short period binary orbit of the LMC and SMC. 

In our models, we do not find conclusive evidence of a coherent periodicity imprinted on the simulated tails. The velocity range of the Stream is remarkably well constrained, such that any such periodic amplitude lies on the order of 5 \km{}, as opposed to the $\sim$15 \km{} espoused by Nidever et al. (2008). Dispersive effects from hot halo interactions would expand this range, but simultaneously diffuse extant periodicity too. We cannot discount however the possibility that this disparity is due to projection effects; our model matches the present on-sky inclination, but may be inconsistent at earlier epochs in the Stream formation. Our simulations are also limited by the modelling of a single Cloud in live form, such that the LMC-sourced tail is perturbed by MW/SMC point potentials only. In practice, the mass-dominant (by factor 10) SMC-sourced tail would impose upon the dynamics of the LMC-sourced tail, and likely draw it towards the location of the observed metal-rich filament.

%%%%%%%%%%%%%%%%%%%%%%%%%%%%%%%%%%%%%%%%%%%%%%%%%%%%%%%%%%%%%%%%%%%%%%%%%%%%%%%
\subsection{SMC}

%%%%%%%%%%%%%%%%%%%%%%%%%%%%%%%%%%%%%%%%%%%%%%%%%%%%%%%%%%%%%%%%%%%%%%%%%%%%%%%
\subsubsection{Star formation history of the SMC}

%%%%%%%%%%%%%%%%%%%%%%%%%%%%%%%%%%%%%%%%%%%%%%%%%%%%%%%%%%%%%%%%%%%%%%%%%%%%%%%
\begin{figure}
\includegraphics[width=1.\columnwidth]{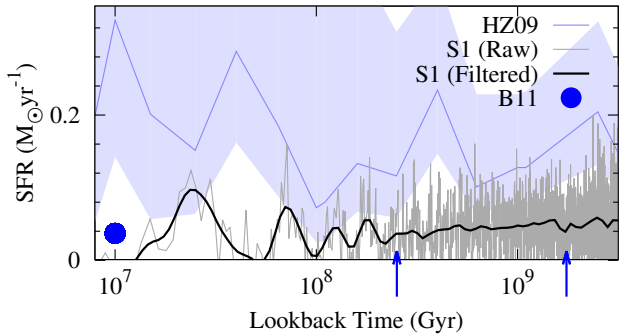}
\caption{The Star Formation History for SMC model S1 (black line) in logarithmic scale of lookback time, in filtered form qualitatively similar to the reduced resolution with lookback time in observational studies. The mean SFH (and dispersion) of the SMC (summed from various subregions; Harris \& Zaritsky 2004) is conveyed with the blue line (shaded region); the current SFR obtained from the extinction-corrected \ha{} map of Bolatto et al. (2011) is indicated with the blue circle. The epochs of close interaction (0.25 and 1.75 Gyr ago, Figure 1) are also highlighted (blue arrows).} 
\end{figure}

Figure 13 compares the simulated SFH of S1 with the observed SFH of Harris \& Zaritksy (2004), with lookback time in logarithmic scale. Our simulation coincides with the epoch in which Weisz et al. (2013) propose an overall enhancement in the SFH compared with its prior queiscent state. Harris \& Zaritsky (2009) adopt a synthetic CMD method to estimate a mean SFR of 0.1 \md\yr{} with a recent enhancement to 0.2 \md\yr; FIR emission (Wilke et al. 2004) and extinction corrected \ha{} (Bolatto et al. 2011) suggest however a lower current global star formation rate of 0.037 to 0.05 \md\yr, which is more consistent with both our model and the expectation that half the presently estimated stellar mass (\ex{3}{8}{} \md; Stanimirovic et al. 2004) is formed in the previous 3 to 4 Gyr (Weisz et al. 2013). Our model successfully maintains a low SFR and correspondingly long gas depletion rate (estimated to be on the order of a Hubble time; Bolatto et al. 2011), in spite of a high gas mass fraction.

The SFH of Harris \& Zaritsky (2009) shows significant long-term enhancements (by a factor 2 to 3) in the SFR more than 1 Gyr ago that are not reproduced by our model. These are consistent with tidal triggering during orbital pericentres (Bergvall et al. 2003), but the resolution of the CMD method decays with lookback time, with the implication that periods of enhanced SFR, persisting on timescales of the order of 100 Myr (Di Matteo et al. 2008), can be poorly resolved. The observational data is also beset by large uncertainty following the wide dispersion in extinction values of younger stars, which have a greater tendency to lie within their dusty formation environments. 

The occurence of episodic bursts are supported however by cluster ages (i.e. Piatti et al. 2005; Glatt et al. 2010). Our model shows evidence of such bursts (albeit weaker), which lag the orbital pericentres by up to several 100 Myr; moreover, each enhancement is preceeded by a drop in SFR. In the context of our SF model, this highlights 1) the delay in tidally-torqued gas infall to form dense clouds; 2) the delay imposed by the chemical conversions and cooling rates within the dense clouds that faciliate SF; and 3) the capacity for SNe (whose progenitor stars live on the order of only $\sim$10 Myr) to destroy dust/\htw{} and weaken SF within the same triggered event. 

%%%%%%%%%%%%%%%%%%%%%%%%%%%%%%%%%%%%%%%%%%%%%%%%%%%%%%%%%%%%%%%%%%%%%%%%%%%%%%%
\subsubsection{Age-Metallicity Relation of the SMC}

%%%%%%%%%%%%%%%%%%%%%%%%%%%%%%%%%%%%%%%%%%%%%%%%%%%%%%%%%%%%%%%%%%%%%%%%%%%%%%%
\begin{figure}
\includegraphics[width=1.\columnwidth]{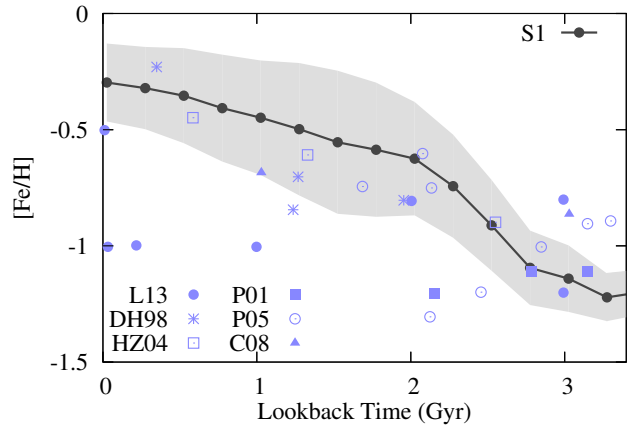}
\caption{The simulated Age-Metallicity Relation for the parent disc of S1 (thick line with 1$\sigma$ shaded), calculated for all new stars (age $<$ 3.4 Gyr, for which chemical evolution is traced). Observed data correspond to the measured clusters of Livanou et al. (2013), Da Costa \& Hatzidimitriou (1998), Harris \& Zaritksky (2004), Piatti et al. (2001, 2005) and Carrera et al. (2008a).} 
\end{figure}

The modest enrichment, punctuated with only short starbursts, is further reflected in the simulated AMR of the parent disc. Figure 14 shows that the new (age $<$ 3.4 Gyr) subpopulation, for which chemical evolution is exclusively traced in our code, overlap with the metal-rich contingent of observed clusters (Da Costa \& Hatzidimitriou 1998; Piatti et al. 2001, 2005; Harris \& Zaritksy 2004; Carrera et al. 2008a; Livanou et al. 2013). In spite of a mean $[Fe/H]\sim-0.7$ (Bernard et al. 2008), De Propis et al. (2010) also find super-solar metal abundances, with which the spread of our data is commensurate; the substantial jump in metallicity among clusters at \ex{6}{8}{} yr ago (i.e. Livanou et al. 2013) is however not reproduced from this subpopulation. 

%%%%%%%%%%%%%%%%%%%%%%%%%%%%%%%%%%%%%%%%%%%%%%%%%%%%%%%%%%%%%%%%%%%%%%%%%%%%%%%
\subsubsection{Spatial distributions of metals and dust in the SMC}

The present-time metallicity profile of S1 exhibits a shallow gradient (Figure 15). The young RGB sample of Carrera et al. (2008a) (adjusted by Cioni 2009, blue dashed line) is well replicated by our outer disc; an older population of AGBs shows no gradient up to 12 kpc (Cioni 2009). The lack of a metallicity gradient beyond radii of 3 kpc is consistent with clusters (Parisi et al. 2009; Da Costa \& Hatzidimitriou 1998), and is continuous with the neighbouring Bridge (Lehner et al. 2008). The SMC is subject to several mechanisms facilitating this flattening, including gas inflow from tidal interactions (Kewley et al. 2010). Strong stellar bars can also efficiently mix the disc ISM, but the \ha{} emission of the SMC bar shows minimal offset from the bar major axis (i.e. Bolatto et al. 2011), indicative of a weak bar (Sheth et al. 2002).

The enhancement towards the inner disc is concordant with the Cepheid distribution, with ages between 30 to 300 Myr, which is closely correlated with recent star formation along the SMC bar (Haschke et al. 2012a). Carrera et al. (2008a) also determine, from the spectra of RGs, that the youngest and most metal-rich stars appear to be concentrated towards the centre, but other studies claim no such trend (Cioni 2009). 

Figure 16 shows that the radial gradient of D traces the stellar metallicity gradient (Figure 15), and is consistent with the observed 1:700 dust-to-gas ratio (Leroy et al. 2007). Furthermore, the factor $\sim5$ variation in D across the SMC bar and inner disc ($r<2$ kpc) reproduces the observed gradient through a slice of the SMC by Stanimirovic et al. (2000). Molecular hydrogen is concentrated homogenously through the stellar bar, similar to barred spirals (Young \& Scoville 1991), with a secondary concentration at $\sim$4 kpc corresponding to the Bridge. The distribution is qualitatively different however to dust surface density and GMCs identified by extended FIR in the SMC (Leroy et al. 2007), which instead manifest in two major sites at opposite ends of the bar. This highlights either dynamical or possibly destructive mechanisms unaccounted for in our model. 

%%%%%%%%%%%%%%%%%%%%%%%%%%%%%%%%%%%%%%%%%%%%%%%%%%%%%%%%%%%%%%%%%%%%%%%%%%%%%%%
\begin{figure}
\includegraphics[width=1.\columnwidth]{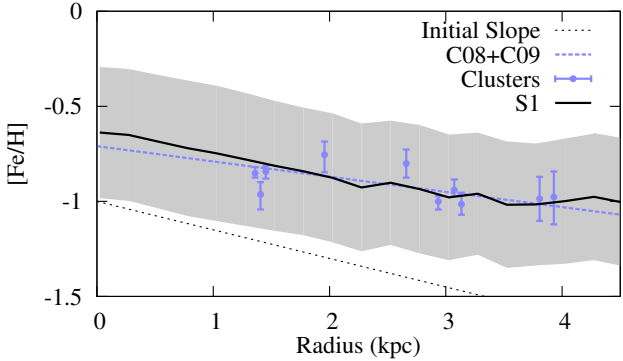}
\caption{Mass-weighted [Fe/H] of new (age $<$ 3.4 Gyr) stars as a function of radius from the photometric centre of SMC model S1. The data are compared with observed [Fe/H] for stellar clusters from Parisi et al. (2009) and Da Costa \& Hatzidimitriou (1998) with corrections by Cioni (2009). The observed best-fit line for youg RGB stars is conveyed with the dotted blue line (Carrera et al. 2008a; Cioni 2009). The initial metallicity slope (-0.15 dex kpc$^{\rm -1}$) used for the model is shown for comparison with the dotted black line.} 
\end{figure}

%%%%%%%%%%%%%%%%%%%%%%%%%%%%%%%%%%%%%%%%%%%%%%%%%%%%%%%%%%%%%%%%%%%%%%%%%%%%%%%
\begin{figure}
\includegraphics[width=1.\columnwidth]{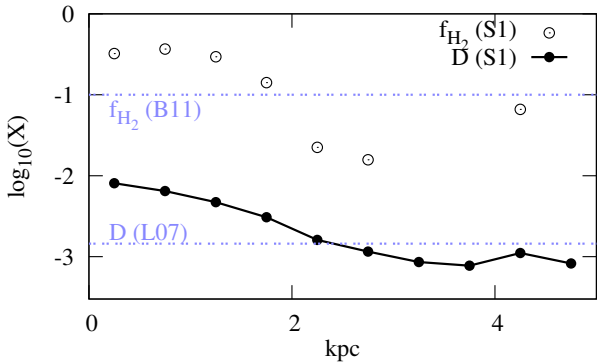}
\caption{Mass-weighted mean radial variations of D and molecular hydrogen fraction (\mfh), with respect to the photometric centre of the SMC model S1 model, For comparison, the SMC shows \mfh{}$\sim$10 per cent (Israel 1997; Bolatto et al. 2011); and the 1:700 dust-to-gas ratio for the SMC (Leroy et al. 2007).} 
\end{figure}

%%%%%%%%%%%%%%%%%%%%%%%%%%%%%%%%%%%%%%%%%%%%%%%%%%%%%%%%%%%%%%%%%%%%%%%%%%%%%%%
\subsection{LMC}

%%%%%%%%%%%%%%%%%%%%%%%%%%%%%%%%%%%%%%%%%%%%%%%%%%%%%%%%%%%%%%%%%%%%%%%%%%%%%%%
\subsubsection{ISM of the LMC}

Figures 5 and 7 indicate that the \hi{}, \htw{} and dust budget of our reference LMC model L1 compares well with recent observations, when utilising the same dust model parameters as S1 (Table 2). The simulated dust mass across the disc matches the observed \ex{1.6}{6}{} \md{}, estimated by Matsuura et al. (2009) from the \hi{} mass and an assumed extinction ratio (Gordon et al. 2003) and D of 1:500. Zhukovska \& Henning (2013) establish a similar range of \ex{1.1}{6}{} to \ex{2.5}{6}{} \md. The overall molecular hydrogen fraction is similar to the estimated ranges from CO and FUV emission (5 to 10 per cent; Israel 1997; Fukui et al. 2008), and thus consistent with only limited destruction of CO clumps under the local ambient field in the LMC (Pineda et al. 2009). Bernard et al. (2008) propose however an IR excess in the LMC is associated with an as yet unseen cold gas component. The \htw{} abundance of GMCs, based on the assumption of virial equilibrium, may also be underestimated given the highly perturbed nature of the LMC.

%%%%%%%%%%%%%%%%%%%%%%%%%%%%%%%%%%%%%%%%%%%%%%%%%%%%%%%%%%%%%%%%%%%%%%%%%%%%%%%
\subsubsection{Tidal influence on stellar morphology and star formation}

As an exemplary Magellanic-Irregular, the LMCs stellar morphology is highly asymmetric/lopsided. Numerical simulations of LMC-analogues indicate that the off-centre bar (with respect to the disc photometric centre), single spiral arm, off-planar warp and, notably, one-sided SF (with respect to the stellar bar), can be the product of short tidal encounters (Besla et al. 2012; Yozin \& Bekki 2014). In this study, we similarly find that the {\it lopsidedness} of our LMC models L1 and L2 (the mean amplitude of the first periodic azimuthal mode between 1.5 and 2.5 disc scalelengths, \asym{I}) exceeds the average for late-type spirals (\asym{I}$=$0.1; Rudnick \& Rix 1998) by a factor of 2 during the near-synchronous pericentres in the LMC/SMC and LMC/MW orbital motions (Figure 1). Moreover, the ellipticity at present time is consistent with observations ($\sim$0.3; van der Marel 2001). 

These distortions are shortlived ($\sim$1 Gyr) and coincident with starbursts. The orbital motion of our SMC model is retrograde with respect to the spin of the LMC disc, which Yozin \& Bekki (2014) show can induce a significant non-symmetric distribution of star forming sites along the major axis of the stellar bar (i.e. one-sided SF), the consequence of infalling gas flows colliding with gas bound to the resonant stellar orbits in the off-centred bar. Our resolutions do not presently have the resolution to model molecular complexes such as 30 Doradus which, lying at one bar end, is responsible for much of the LMCs current star formation rate (Kim et al. 2007; Harris \& Zaritsky 2009). We show in Section 3.3.4, however, that dust and \htw{} in L1 similarly show non-axisymmetry in their distributions, with an analogue to the {\it Molecular ridge}, a series of \htw{} clumps in proximity to 30 Doradus (Ott et al. 2008).

We note that for both LMC models, the SFR of the recent 3 Gyr is systematically less than the 0.1 to 0.4 \md\yr{} calculated by Harris \& Zaritsky (2009) by a factor of 2 to 3, while the present time SFR lies towards the lower bound on other estimates ($\sim$0.06 to 0.26 \md\yr; Calzetti et al. 2007; Kennicutt et al. 1995). We ascribe this discrepancy to numerically unresolved SF regions that dominate the observational estimates, and poor constraints on the initial gas mass of the LMC in our simulations.

%%%%%%%%%%%%%%%%%%%%%%%%%%%%%%%%%%%%%%%%%%%%%%%%%%%%%%%%%%%%%%%%%%%%%%%%%%%%%%%
%\begin{figure}
%\includegraphics[width=1.\columnwidth]{f18.eps}
%\caption{The Star Formation History for LMC model L2 in logarithmic scale of lookback time (red line). An isolated model (L2-Isolated) is also shown (blue line). The current SFR obtained from \ha{} and IR emission {\bf (Calzetti et al. 2007; C07)} and free-free flux (Murray \& Rahman 2010; MR10) are indicated with the blue {\bf circle and triangle respectively}. The epochs of close interaction (0.3 and 1.7 Gyr ago, Figure 1) are also highlighted (blue arrows).}
%\end{figure}

We emphasise however the agreement of our model to within the uncertainties of the overall \htw{}/dust mass, and the AMR. Figure 17 shows the mass-weighted AMR for LMC models, compared to clusters (Harris \& Zaritsky 2009; Livanou et al. 2013), and deep NIR surveys (Rubele et al. 2012). All models show similarly modest enrichment until 2 Gyr ago; this is retained by the high mass LMC model (L1), which reproduces the well constrained metallicities within the recent 0.5 Gyr. The apparent {\it jump} in recent stellar metallicity inferred by Livanou et al. (2013) and commencing from \ex{6}{8}{} yr ago can be invoked in the massive LMC model with a more massive SMC (M$_{\rm SMC}\times$2), such that the 1$\sigma$ spread can even extend to observed solar to super-solar metallicities (Livanou et al. 2013). The observed metallicities at 1.7 to 2 Gyr ago convey a wide dispersion, encompassing the metallicities of the 0.1 \zd{} Stream (Fox et al. 2013) and metal-rich filament (Richter et al. 2013). The data of Rubele et al. (2012) imply however a large metallicity gradient at this epoch, such that the outer disc, with the greater proclivity for tidal stripping, would not account for the [S/H]$\sim$-0.3) \hi{} filament.

%%%%%%%%%%%%%%%%%%%%%%%%%%%%%%%%%%%%%%%%%%%%%%%%%%%%%%%%%%%%%%%%%%%%%%%%%%%%%%%
\begin{figure}
\includegraphics[width=1.\columnwidth]{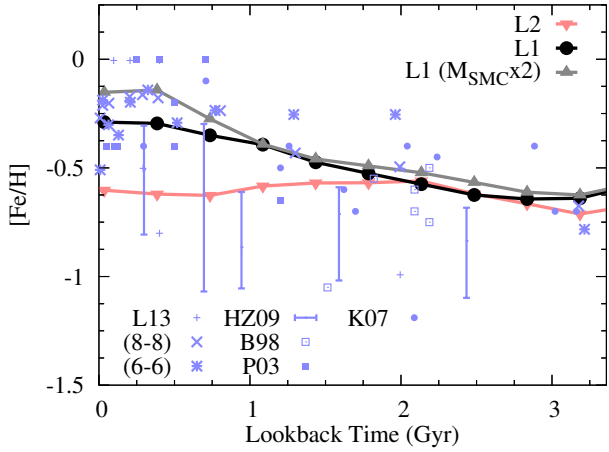}
\caption{The Age-Metallicity Relation, for new stars (age $<$ 3.4 Gyr) of LMC models L1, L2, and an L1 model differing by the SMC mass alone (M$_{\rm SMC}\times$2). The data are compared with cluster data from Livanou et al. (2013), Harris \& Zaritksy (2009), Bica et al. (1998), Piatti et al. (2003) and Kerber et al. 2007). The averaged metallicity from NIR tiles (8-8) and (6-6) from Rubele et al. (2012) are also shown.} 
\end{figure}

The lower mass LMC model L2, whose tidal triggered SF is stronger, undergoes net {\it de-enrichment} commencing at this epoch, coincident with a decline in dust mass; conversely, the \htw{} abundance increases rapidly. This dichotomy is a consequence of tidally-torqued \hi{} infalling to form substantial dense \htw{} clumps that, while effective in forming new stars, concurrently destroy extant dust by a concentration of new SNe in the inner disc. This de-enrichment is not observed in the LMC, nor are such nuclear concentrations of \htw{} (Section 3.3.4).

%%%%%%%%%%%%%%%%%%%%%%%%%%%%%%%%%%%%%%%%%%%%%%%%%%%%%%%%%%%%%%%%%%%%%%%%%%%%%%%
\begin{figure}
\includegraphics[width=1.\columnwidth]{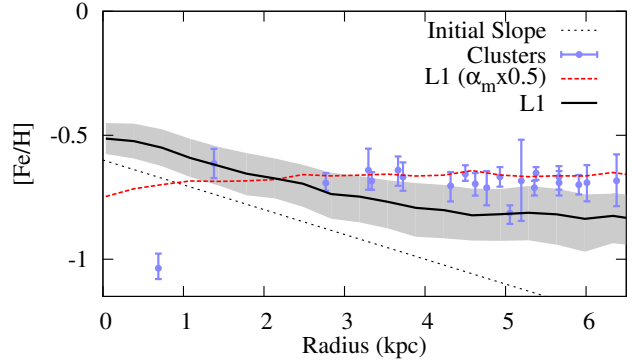}
\caption{Metallicity Gradient of new stars (age $<$ 3.4 Gyr) in the LMC model L1 at zero lookback time (thick black line, with 1$\sigma$ dispersion in the shaded region), with Iron abundance as a function of radial distance from the LMC disc centre. The initial metallicity slope is indicated with the dotted black line. An L1 model differing by initial slope alone ($\alpha_{\rm m}$$\times$0.5) is shown with the dashed red line. The results are compared with clusters from Grocholski et al. (2006), which indiate a flat profile beyond the bar radius of 1.5 kpc (van der Marel et al. 2002).}
\end{figure}

%%%%%%%%%%%%%%%%%%%%%%%%%%%%%%%%%%%%%%%%%%%%%%%%%%%%%%%%%%%%%%%%%%%%%%%%%%%%%%%
\subsubsection{Spatial distributions in the LMC}

The outer metallicity profile of LMC model L1 (Figure 18) flattens during its recent evolution (relative to the initial gradient $\alpha_{\rm m}$$=$-0.1), consistent with the present abundance distribution of RR Lyrae (Haschke et al. 2012b) and clusters (e.g. Grocholski et al. 2006). The stellar bar has been considered responsible for this flattening through the mixing of stellar populations (Sharma et al. 2009). Moreover, its effectiveness in driving gas flows since its apparently recent triggered formation is indicated by an estimated 35 per cent of stellar material in the bar being $\le$3 Gyr old (Smecker-Hane et al. 2002). 

The simulated dust-to-gas ratio profile of L1 is more resilient to this flattening and retains a distinct gradient. Bernard et al. (2008) and Dobashi et al. (2008) similarly find observational evidence of a dust gradient, with the latter noting an increase of the extinction-to-column density (A$_{\epsilon}$/N$_{\rm H}$) from the outer regions of the LMC to the 30 Doradus complex, which is interpreted as an increase in dust abundance close to star formation sites {\it or} a CO-dark \htw{} component. The simulated dust traces the location of the bar; similarly, observed PAHs appear to be enhanced along the LMC stellar bar (Paradis et al. 2009; Meixner et al. 2010). Kim et al. (2010) note that D is relatively uniform across the LMC, except in the vicinity of supershells, which are clustered around the bar (Williams et al. 1999).

Molecular clouds in the LMC are distinct and well spread throughout the disc (Fukui et al. 2008), and Kawamura et al. (2009) establish the physical association between very young clusters in the LMC and GMCs. Figure 19 shows that \htw{} clumps are widespread in our L1 simulation, but predominantly lie in the vicinity of dust. A stronger spatial correlation could be claimed however with dynamical features such as the central bar and a major tidal arm. Furthermore, a substructure analogous to the continuous ridge of CO emission south of 30 Doradus, called the {\it Molecular Ridge} (e.g. Ott et al. 2008) extends from one end of the bar. This ridge is not clearly traced by the dust intensity, which is mainly concentrated along the bar major axis. 

%%%%%%%%%%%%%%%%%%%%%%%%%%%%%%%%%%%%%%%%%%%%%%%%%%%%%%%%%%%%%%%%%%%%%%%%%%%%%%%
\subsubsection{Dependence of LMC evolution on dust and metal parameters}

The robustness of the massive LMC model (unlike the highly tidally-disturbed SMC model) permits us to explore how its evolution depends on dust parametrisations. The most striking result, shown in Figure 20, is the substantial {\it de-enrichment} that follows from (i) halving the reference destruction timescale \tdesto{}, such that it is equivalent to \tacco{} (\tdesto$\times$0.5); and (ii) halving both \tdesto{} and \tacco{} (\tacco$\times$0.5). Increasing both \tdesto{} and \tacco{} (\tacco$\times$2) yields the opposite evolutionary behaviour. These trends are mirrored in the global dust mass evolution (i.e. a reduced \tdesto{} leads to systematic decline in dust mass), highlighting the coupling of stellar metalicity to dust abundance. 

The similarity in the AMRs corresponding to aforementioned scenarios (i) and (ii) implies the AMR is governed more by the choice of \tdesto{} than \tacco{}. We argue in Section 4.1.2 that \tdest{} in our dust model is a proxy for the local SN rate, and thus the dependence on \tdesto{} reflects the common observation that dust mass is constrained by SNe, as both major sources and sinks of dust. Accordingly, the model is also sensitive to the timescale over which SNe expands adiabatically into the ISM (i.e \tsn{}$\times$2).

In an alternative model where the initial abundance slope is halved ($\alpha_{\rm m}$$=$-0.05 dex kpc$^{\rm -1}$) we find a 0.4 dex {\it dip} in the AMR $\sim$1.5 Gyr ago. A shallower dip is found in the observed AMR (Harris \& Zaritksy 2009), which Bekki \& Tsumjimoto (2012) attribute to the transfer of at least $\sim$10$^{\rm 9}$\md{} metal-poor gas from the SMC. While we do not model external gas infall on our live model, this mass value cannot be fully reconciled with that stripped from the S1 model (Figure 5). Instead, we find the dip can be traced to the anomalously metal-poor inner disc (Figure 20), which we ascribe to a combination of 1) highly efficient early SF in the inner disc destroys extant dust, inhibiting later nuclear SF; 2) efficient early SF enhances disc self-gravity, leading to a stronger bar that dilutes the inner disc by drawing in metal-poor gas from the outer disc (i.e. Friedl \& Benz 1995); and 3) highly efficient dust accretion in the metal gas-phase promotes \htw{} formation and SF, but leaves only depleted gas for subsequent astration. Exploring the coupling of these mechanisms would serve a useful topic for future work.

%%%%%%%%%%%%%%%%%%%%%%%%%%%%%%%%%%%%%%%%%%%%%%%%%%%%%%%%%%%%%%%%%%%%%%%%%%%%%%%
\begin{figure}
\includegraphics[width=1.\columnwidth]{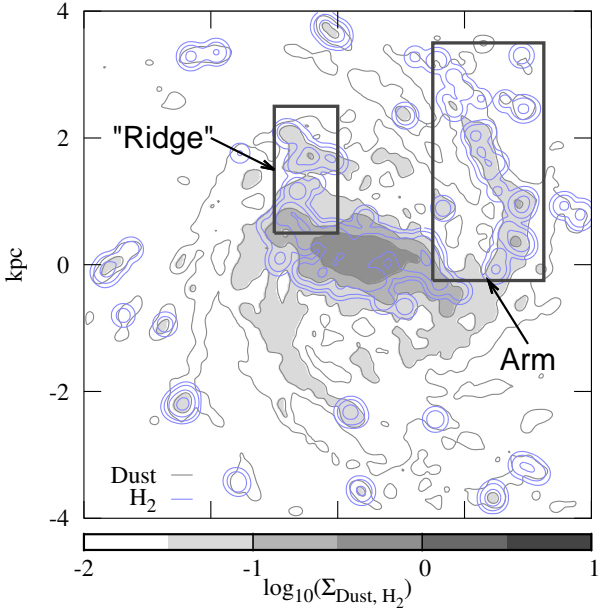}
\caption{Surface mass density (\md{}pc$^{\rm -2}$) of dust (black contours/grey colour scale) and \htw{} (blue contours) with logarithmic scaling. The stellar bar and a single prominent tidal arm are clearly traced by both mediums. We also highlight a possible analogue to the Molecular Ridge which is similarly quasi-coherent, and is not a spiral structure despite protruding from a single bar end.} 
\end{figure}

%%%%%%%%%%%%%%%%%%%%%%%%%%%%%%%%%%%%%%%%%%%%%%%%%%%%%%%%%%%%%%%%%%%%%%%%%%%%%%%
\begin{figure}
\includegraphics[width=1.\columnwidth]{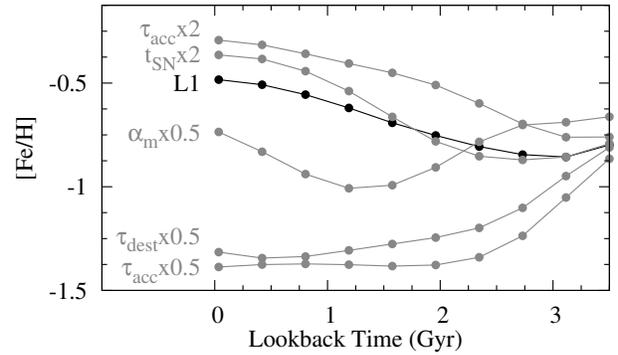}
\caption{The parameter dependence of the AMR. Reference model L1 is compared with models with a shallower initial metallicity gradient ($\alpha_{\rm m}$$\times$0.5$=$-0.05 dex kpc$^{\rm -1}$), increase SNe adiabatic timescale (\tsn{}$\times$2), increased and decreased dust accretion timescale (\tacco$\times2$=\ex{5}{7}{} yr, \tacco$\times0.5=$\ex{1.25}{7}{} yr respectively) and equal dust accretion and destruction timescales (\tdest$\times0.5=$\ex{2.5}{7}{} yr).} 
\end{figure}

%%%%%%%%%%%%%%%%%%%%%%%%%%%%%%%%%%%%%%%%%%%%%%%%%%%%%%%%%%%%%%%%%%%%%%%%%%%%%%%
\subsection{Composition of the Bridge}

Figure 21 shows properties of the simulated Bridge across its $\sim$20 degree extent from the combined data of S1 and L1. The LMC model L1 contributes material to this \hi-structure, consistent with the observed bimodality (Muller et al. 2003), and unlike its purely-tidal role in shaping the simulated Stream. This bi-directional stripping results in a typical density of around the observed \ex{5}{20}{} \cms{} (Br\'{u}ns et al. 2005). Since we model the LMC and SMC as separate live models, we can only speculate as to whether shock fronts at these converging \hi{} flows collide and collapse to form \htw{} clumps. Nonetheless, young stars (age less than 300 Myr) are present throughout the Bridge (Figure 22), as found by Mizuno et al. (2006).
 
We find a generally low velocity mean dispersion along the Bridge, consistent with the observed 20 kms$^{\rm -1}$ (Br\'{u}ns et al. 2004). At smaller scales, Mizuno et al. (2006) find localised dispersions for CO-emission sites as low as 2 kms$^{\rm -1}$, implying {\it in-situ} formation rather than removal from the parent MC. In our simulation, the low gravitational stability implied in the high density analogue to the SMC {\it tail} coincides with the largest concentration of \htw{}, dust and new stars (with lifetime shorter than the 300 Myr since the previous close encounter with the LMC). The peak 40 kms$^{\rm -1}$ towards the centre of the Bridge results from multiply different kinematic components (such as the {\it counter-bridge}) lying along this sightline, similarly found by Br\'{u}ns et al. (2004).

The dust-to-gas ratio in the tail (4 to 8$^{\circ}$ along the SMC-to-LMC path) agrees well with Gordon et al. (2009). This is lower than that expected from a metallicity-dependence, leading the authors to propose ongoing dust destruction. This is concordant with the abundant young stellar concentration in the simulation, and acknowledged RSG, OB stars and \hi{} shells in the SMC tail (Irwin, Kunkel \& Demers 1985; Muller et al. 2003; Boyer et al. 2011; Harris 2007). The average molecular hydrogen fraction of the simulated tail (and the Bridge as a whole) exceeds the 0.2 per cent estimated from CO-emission (Mizuno et al. 2006) by a magnitude, in spite of their using a X$_{\rm CO}$ appropriate for the 0.1 \zd{} SMC. This excess mass may instead exist in ionized form, which Lehner et al. (2008) propose accounts for 90 per cent of the Bridge. 

The metallicity of the Bridge is contiguous with the SMC metallicity gradient (Figure 15), implying its stripping from the outer gas disc. The [Fe/H] of new stars (age less than 3.4 Gyr) averages to $\sim$0.1 \zd; the ISM is 0.2 dex more enriched. Our results are consistent with previous simulations of the Bridge with a more simplified approach to chemical evolution (Bekki \& Chiba 2005; 2007), but marginally exceed the gas-phase abundance established from sightlines (Lehner et al. 2001; 2008). An apparently large variation in metallicity between the wing and bridge, as inferred from Lee et al. (2005) and Lehner et al. (2008), cannot be accounted for in our models. We note however the wide spread in observed abundance measurements (ISM, Lehner et al. 2009; B-type stars, Dufton et al. 2008; clusters, Livanou et al. 2013), which hinders efforts to constrain our model parameters at this time.

%%%%%%%%%%%%%%%%%%%%%%%%%%%%%%%%%%%%%%%%%%%%%%%%%%%%%%%%
\section{Discussion}

%%%%%%%%%%%%%%%%%%%%%%%%%%%%%%%%%%%%%%%%%%%%%%%%%%%%%%%%%%%%%%%%%%%%%%%%%%%%%%%
\begin{figure}
\includegraphics[width=1.\columnwidth]{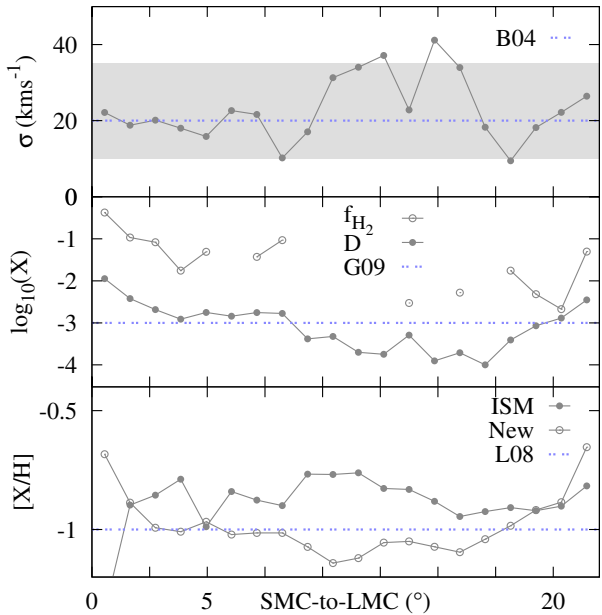}
\caption{(Top panel) Second moment of radial velocity in the Bridge as a function of position (in degrees) from the SMC to LMC. The mean (1$\sigma$) from observed \hi{} spectra (Br\'{u}ns et al. 2004) is shown with the blue dashed line (shaded region). The {\it tail} region spans approximately from 3.5 to 10 degrees; (Middle panel) mass fraction of Molecular Hydrogen (\mfh{}) and the Dust-to-Gas ratio (with observed D from Gordon et al. 2009); (Bottom panel) mean metallicity of new stars (dotted black line) and gas-phase ISM (thick black line).}
\end{figure}

Table 3 provides a cursory diagnosis of our simulations' capacity to address the myriad observations outlined in Table 1. The present study demonstrates that our ISM model and the tidal-dominated paradigm for the current morphology can qualitatively reproduce the recent evolution of the MCs. We discuss in the following section caveats and improvements for our models.

%%%%%%%%%%%%%%%%%%%%%%%%%%%%%%%%%%%%%%%%%%%%%%%%%%%%%%%%%%%%%%%%%%%%%%%%%%%%%%%
\begin{table*}
\centering
%\begin{minipage}[t]{1.\columnwidth}
\caption{Comparison of observations outlined in Table 1 with the scope of results attained by our simulations}
\begin{tabular}{@{}ccc@{}}
& Property & Reproduced? \\
\hline
Stream & \hi{} mass/column density & Matches morphology, mass deficit of factor $\sim$5 \\ 
& Gas-Phase Metallicity & Yes \\
& Depletion & No, unless depletion very efficient prior to tidal stripping \\
& Gas-to-Dust ratio & Consistent with upper observable limit \\ 
& Maximum stellar density & Yes, for low mass LMC and SMC-sourced streams \\
\hline
Bridge/Wing & Gas-Phase metallicity & Within 0.2 dex of observation \\
& Stellar Metallicity & Yes \\
& \htw{} mass fraction & No, simulation estimates more \htw{} \\
& Dust-to-Gas Ratio & Yes, in {\it wing} region \\
\hline
SMC & SFH & Matches gas depletion timescale, but weak starbursts \\
& Current SFR & Yes \\ 
& AMR & Yes \\
& Metallicity Gradient & Yes \\
& \htw{} mass & Yes, but concentrated in bar \\
& Dust mass & Yes, but concentrated in bar \\
\hline
LMC & SFH & Consistent with observed lower limits \\
& Age-Metallicity Relation & Yes \\
& Metallicity Gradient & Yes \\
& \htw{} mass (\md) & Yes, but too centrally concentrated \\
& Dust mass (\md) & Yes, but too centrally concentrated \\
\hline
\end{tabular}
%\end{minipage}
\end{table*}

%%%%%%%%%%%%%%%%%%%%%%% 4.1
\subsection{Dust abundance as a function of metallicity and environment}

%%%%%%%%%%%%%%%%%%%%%%%%%%%%%%%%%%%%%%%%%%%%%%%%%%%%%%%%%%%%%%%%%%%%%%%%%%%%%%%
\begin{figure}
\includegraphics[width=1.\columnwidth]{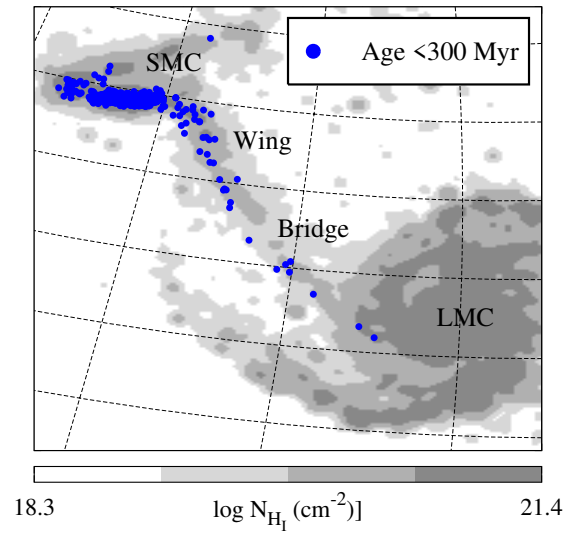}
\caption{\hi{} contour map, plotted in the equatorial coordinate system similar to Putman et al. (1998), centred on the simulated Bridge (combining S1 and L1 simulation data at zero lookback time) with scale limits as in Figure 4. The blue dots represent stellar particles from the SMC with age less than 300 Myr, showing a significant presence across the {\it tail} region and up to the LMC disc.}
\end{figure}

%1 to first order, d-Z
%2 our models, compare l,s,mw, fornax
%3 dust source, carbon output not strongly dependent on Z, but absent 2175 in SMC, due to less self-shielding from silicates
%4 does deficit arise from destruction, poor accretion
%5 metal loss

To first-order, the global dust-to-gas ratios across our simulated LMC and SMC follow a linear correlation with their respective mean metal abundances (Figure 23), qualitatively matching previous numerical work (B13) and an observed correlation in observational samples (i.e. Dwek 1998; Zafar \& Wilson 2013). This dust-to-metal ratio (\rdm) is maintained by the model self-consistently (i.e. not explicitly set as a constraint, except as an initial value) as shown in the D-\ao{} phase space (Figure 24) This contrasts with present observational estimates of D in the SMC, which are a factor of 5 to 14 less than the Galactic ratio (Sofia et al. 2006; Leroy et al. 2007; Gordon et al. 2009), and thus imply a deficit in dust when compared to the factor 4 to 5 expected from differences in mean metallicity. 

Our calculation of \rdm{} from simulations are free of line-of-sight uncertainties and selection bias towards warm dust. The line-of-sight morphology of the SMC is complex where, besides tidal effects (Nidever et al. 2011; Diaz \& Bekki 2012), feedback within the ostensibly weak potential of the SMC influences vertical structure and therefore measures of attenuation. This also makes a representative CO-\htw{} conversion factor for the SMC difficult to ascertain (Leroy et al. 2011). The presently low observed D could also be explained by an underestimation of the contribution from cold dust lying towards the limits of equilibrium; Leroy et al. (2007) derive an upper constraint on the dust mass at $\sim$2 times the 10$^{\rm 6}$ \md{} mass commonly attributed to the SMC. Galametz et al. (2011) show that sub-millimetre data can be combined with SED analysis in low-metallicity dwarves to recover the universal \rdm{}. 

In practice, dwarf galaxies exhibit wide dispersion in \rdm{}, which various studies attribute to the variability in dust sources with metallicity, stellar outflows, SFH and destruction efficiencies (Lisenfeld \& Ferrara 1998; Mattsson et al. 2014). Furthermore, tidal perturbations disrupt the potential and drive gas dynamics, upon which the ISM composition depends. Mergers can reduce \rdm{} following the triggered consumption of metal-rich gas (Hayward et al. 2011), and tidal stripping appears to preferentially remove the metal-enriched \hi{} over dust (Pappalardo et al. 2012). This manifests in our simulated Stream and Leading Arms, which show negligible dust content (Figure 23). However, observational estimates for D in the Stream filaments (coincident with sightlines Fairall 9 and RBS 144; Fox et al. 2013) far exceed that in our reference model by orders of magnitude (Figure 8). In the following sections, we attempt to reconcile our results with the broader discussion of interstellar dust evolution.

% argue that adoption of high SNe efficiency leads to an efficient dust growth phase that obscures sensitivity to
% initial Rdm; SNe in this case cannot constrain dust growth
% accretion dominant thereafter
% destruction timescale more important than accretion, consistent with SNe as major constraints on dust
% relative timescale of MCs consistent with SFR

%%%%%%%%%%%%%%%%%%%%%%% 4.1
\subsubsection{Dust synthesis and growth}

%%%%%%%%%%%%%%%%%%%%%%%%%%%%%%%%%%%%%%%%%%%%%%%%%%%%%%%%%%%%%%%%%%%%%%%%%%%%%%%
\begin{figure}
\includegraphics[width=1.\columnwidth]{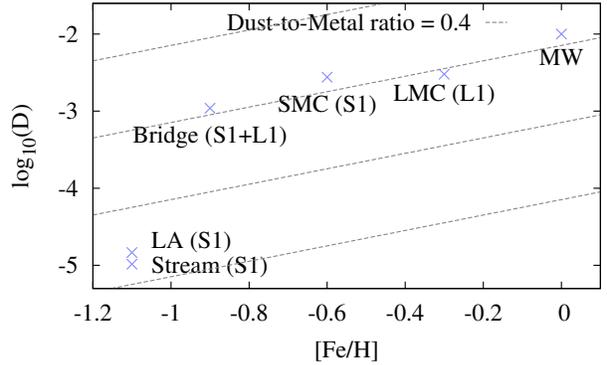}
\caption{Dust-to-gas ratio as a function of mean metallicity for our simulated LMC (model L1), SMC (S1), Bridge (S1+L1), and the Leading Arm and Stream features as formed primarily from stripped SMC (S1) material. The linear relation (proportionality constant of 0.4; dashed lines) proposed by previous works (i.e. Dwek 1998) matches the trend implied by the simulated LMC, SMC, Bridge and the Galactic D (0.01; Knapp \& Kerr 1974) at solar metallicity.} 
\end{figure}

%%%%%%%%%%%%%%%%%%%%%%%%%%%%%%%%%%%%%%%%%%%%%%%%%%%%%%%%%%%%%%%%%%%%%%%%%%%%%%%
%\begin{figure}
%\includegraphics[width=1.\columnwidth]{f25.eps}
%\caption{Total dust mass in the parent disc of LMC and SMC models as a function of time. The reference models L1 and S1 %(with initial dust-to-metal \rdm{}$=$0.4) are compared with similar models with initial ratio \rdm{}$=$0.1.}
%\end{figure}

% stellar sources vary, condensation efficiency high for SNe, counteracts variance in initial dust mass
% accretion is dominant mode thereafter

The contributions of different stellar sources to the metals and dust in the ISM are not well understood for the MW/MCs (Gehrz 1989; Boyer et al. 2011). AGBs convey a variability cycle in mass loss (Srinivasan et al. 2009) and large scatter with metallicity (Boyer et al. 2012), while the dust mass in SN remnants ranges from conservative lower limits of 0.01 \md{} to an estimated $\sim$0.5 \md{} in SN 1987a in the LMC (Matsuura et al. 2011). The latter uncertainty largely prohibits a constraint on dust grain growth in the ISM (Draine 2009; Boyer et al. 2012). The relative delay in dust output from different stellar sources (Dwek 1998) is also a major complication in the MCs, which are presently experiencing starbursts and whose current dust output is compounded by the delayed products from prior starbursts in their very recent evolution (Harris \& Zaritksy 2009).

% C/O 

The evolution in dust sources can manifest in the C/O ratio (Dwek 1998; Zhukovska \& Henning 2013). Given our adoption of simple ejecta yields (Dwek 1998, instead of that from more recent work i.e. Ferrarotti \& Gail 2006), and the tracking of elemental abundances for only the 3.4 Gyr of our simulations, we cannot fully compare our results with more detailed analytical studies (i.e. Zhukovska \& Henning 2013). Nonetheless, we find the C/O ratio drops at commencement of the simulations, consistent with tidally-triggered starbursts that invoke large silicate production in short-lived massive stars. Comparing the final C/O ratios of our MC models with the analytical Galaxy model of Dwek (1998), we find consistency with an anticorrelation of oxygen abundance with metallicity among dIrrs (Garnett et al. 1995). Two observations of the MCs are further consistent with this result. First, the 2175 \AA{} feature is weak in the LMC and absent in the SMC (Gordon et al. 2003) with the implication that small carbonaceous grains are destroyed more efficiently in the harsher FUV background of the SMC. Secondly, Matsuura et al. (2013) find substantial SNe gas feedback in the MCs but more significantly so in the SMC (even when accounting for uncertainty in mass loss rates and the assumed AGB gas-to-dust ratio), in accordance with relatively more enhanced SFR or weaker stellar winds in the more metal-poor environment.

% dust yield from SNe important in wider discussion of the rdm
% correlation in dust and stellar density suggests rapid production and constrained by SNe, high redshift
% Mattsson et al. (2014) suggest that stellar sources and sinks efficient, but accretion varies with metallicity
% agree with Zafar \& Wilson (2013) that rapid enrichment by SNe can maintain the universal ratio
% accretion expected to be inefficient in metal-poor galaxies; filling factor low due to low C0 etc
%Mattsson et al. (2014) argue that dust accretion can balance destruction, and thus dust yields are metallicity dependent, yielding a variant Rdm

The dominant dust sources can be also revealed by constraining the variation of dust mass with metal abundance, since the yields of stellar winds and accretion efficiency are expected to vary with metallicity. For instance, Zafar \& Wilson (2013) interpret a constant \rdm{} (similar to the Galactic value adopted by Dwek 1998) among a recent sample of GRB/QSOs across redshifts 0.1 to 0.63, as evidence for rapid dust production. Their metallicity range encompasses that of the MCs, appearing to support our adopted dust model parametrisations. A spatial correlation of extinction and stellar density (Grootes et al. 2013), and the slow synthesis of dust by AGBs, suggests SNe are responsible for this production rate, as further demonstrated by our self-consistent work, analytical models (Lisenfeld \& Ferrara 1998; Dwek 1998), and other studies attempting to explain high redshift dust masses (Dunne et al. 2003). On the other hand, De Cia et al. (2013) find strong metallicity dependence of \rdm{} among another sample of quasar foreground DLAs. Mattsson et al. (2014) try to reconcile these studies by proposing that while stellar yields vary with metallicity, rates of dust accretion and destruction by SNe will tend towards an equilibrium that manifests in an apparent constant \rdm{}.

% high efficiency obscures dependence on Rdm 

Our models provide some support to the concept of an equilibrium state. Figure 24 compares the evolutionary tracks, within the D-\ao{} phase space, of our reference LMC model L1 with additional models covering a small subset of the ISM model parameter space. Model L1, with an initial \rdm{} set at the Galactic value ($\sim$0.4, Dwek 1998), is self-consistently bound to this \rdm{} throughout its evolution, similar to analytical models (i.e. Edmunds 2001). The bottom panel highlights the substantial sensitivity of our simulations to characteristic timescales of ISM processing, which we conveyed earlier in the AMR (Figure 20). Our model is however robust to variations in the initial \rdm{} (Figure 24, top panel) where, over the 3.4 Gyr duration of the simulation, an initially dust poor model converges upon our reference model in which we had assumed almost complete condensation of gas-phase metals. 

% condensation efficiency

Our results thus far are consistent with efficient dust production by SNe. However, a recent analytical model by Zhukovska (2014) tracks the evolution of low metallicity galaxies in the D-\ao{} plane, from which she argues that the SNe II condensation efficiency is a governing parameter. Hirashita (1999) similarly showed that larger efficiencies correspond almost linearly to higher D. For brevity, we adopted in this study the efficiencies derived by Dwek (1998), which are typically a order larger than those advocated in the cited one-zone studies (i.e. 0.01 to 0.1). It is presently unclear, for the purposes of self-consistent models, an appropriate choice of this parameter; the dust yield from SNe may be invariant over a metallicity range, but the condensation efficiency into its surroundings may depend indirectly on metallicity via cooling rates/shielding (Mattsson et al. 2014). Moreover, elemental depletion observed in the vicinty of SN ejecta could indicate an additional cold dust component unaccounted for in the low efficiencies commonly derived (Morgan \& Edmunds 2003; and references therein).

% missing dust mass problem in lmc matsuura +09 suggest accretion important
% need to refine C/O ratio and efficiency to establish correct balance

In the specific case of the MCs, however, dust output rates of SNe/AGBs are insufficient, by up to magnitude, to account for the global dust mass (Boyer et al. 2012; Zhukovska \& Henning 2013). This is supported by the spatial correlation of the dust surface density with atomic and molecular hydrogen (Roman-Duval et al. 2010; Sandstrom et al. 2010) indicating therefore the dominant role of dust growth in the ISM. By contrast, we find \tacc{} is a weak parameter for our LMC models (Section 3.3.3). We leave the task of reproducing the observed relative roles of dust growth by stellar sources and accretion to a subsequent paper. We generally find that the final dust mass is set by the sensitive balance of: the initial metallicity profile ([m/H]$_{\rm R=0}$, $\alpha_{\rm m}$), the initial dust-to-metal ratio (\rdm{}), reference dust accretion timescale (\tacco{}), the adiabatic timescale of SN (\tsn{}), the stellar dust yield and condensation efficiencies, and the tidal environment which triggers star formation and shapes the distribution of dust/\htw{}. To date, these factors are not well constrained for the metal-poor MCs, and thus we do not attempt further fine-tuning of our model parameters against observational constraints (i.e. Table 1).

\begin{figure}
\includegraphics[width=1.\columnwidth]{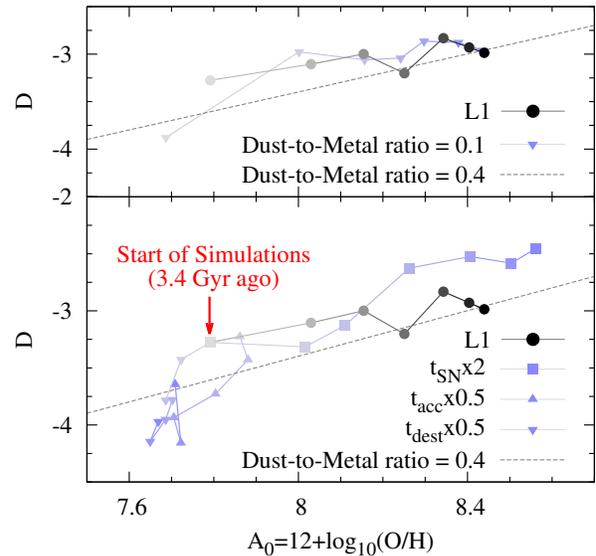}
\caption{(Bottom panel) Dust-to-gas ratio as a function of mean oxygen abundance for LMC model L1, and models which vary by increased adiabatic timescale for SNe (\tsn$\times$2), shorter dust accretion timescale (\tacco$\times$0.5) and a 1:1 ratio in destruction and accretion timescales (\tdest$\times$0.5). The symbol colour tone varies from light (3.4 Gyr ago) to dark (0 Gyr ago), highlighting how reference model L1 evolves with a near constant \rdm{}, whereas deviations to the ISM timescales can yield very different behaviour; (Top) A comparison of L1 and a model with initial \rdm{} of 0.1. Note that both models converge after commencing from different locations in the D-\ao{} phase space.} 
\end{figure}

%%%%%%%%%%%%%%%%%%%%%%%%%%%%%%%%%%%%%%%%%%%%%%%%%%%%%%%%%%%%%%%%%%%%%%%%%%%%%%%
\subsubsection{Destruction timescale}

% a major constraint on dust mass is \tdesto{}
% argue that our model is a proxy for the SN rate
% ratio of tdests peaks with tidal starburst, possibly includes some stripped gas, but mainly reflects ?

Dwek (1998) derives the Galactic accretion and destruction timescales independently; the former depends on the time to accrete a typical layer thickness on dust grains and the mass fraction of molecular clouds; the latter follows from modelling of radiative shocks. These factors were assumed constant over a Hubble time evolution, despite a wide dynamical range in local conditions, leading to the canonical \tdest{}$=$2\tacc{} relation adopted in this work. This is a justifiable simplifying assumption, but the recent evolution of the MCs is clearly not steady. More recent studies have since argued for a decoupling of the timescales, where \tdest{} should directly reflect the local SN rate (i.e. Inoue 2003; Zhukovska \& Henning 2013). 

Our simulations are the first to provide per-particle allocation of \tacc{} (and \tdest{}), established per timestep from the local $\rho$ and thermal speed (Equation 7). While not decoupled, our cursory parameter studies indicate the galaxy-wide AMR depends more sensitively on the parametrized \tdest{} than \tacc{} (Section 3.3.3), supporting our earlier assertion that SNe are significant dust sources in our model. This conflicts with studies indicating the dominance of dust accretion over its production to accomodate the current budget in the MCs (see Section 4.1.1). It is nonetheless instructive to compare destruction efficiencies in our LMC and SMC models.

Observational studies have adopted the SN rate per unit surface area to establish the mean dust lifetime, implicitly assuming a constant SFR over the corresponding timescales. For the LMC, a range of 0.4 to 1.1 Gyr, depending on grain type and subject to the error bounds of the SFR is obtained (Zhukovska \& Henning 2013). By a similar method, Matsuura et al. (2013) estimates lifetimes of $\sim$\ex{0.4}{9}{} yr and $\sim$\ex{1.4}{9}{} yr for the LMC and SMC respectively. These data suggest dust destruction is less efficient in the SMC by an approximate factor of 3.

Our dust model does not explicitly account for microscopic-scale grain sputtering in the diffuse warm gas behind the SN remnant, since simulations cannot at present distinguish this region due to resolution limitations. Instead, dust destruction in our models, as parametrised by \tdest{}, is assumed to be most efficient in proximity to the thermal radiation ejected from SNe emerging from dense clouds (where Equation 7 is imposed only on those gas particles lying within the smoothing length of a given SN). Our parametrized \tdest{} is thus a proxy for a local SN rate dependence, with which we evaluate the global destruction timescale in our reference LMC and SMC models, to compare with these observational estimates. We consider only gas particles lying in the line-of-sight region of the sky for which the integrated SFH histories (a proxy for the SN rate, if combined with the IMF) of the MCs were established by Harris \& Zaritsky (2004; 2009). 

The main panel of Figure 25 shows the PDF of \tdest{} for individual gas particle lifetimes (at zero lookback time), ascertained with Equation 7 and \tdest{}$=$2\tacc{}. The PDF peaks correspond to short \tdest{} on the order of the reference Galactic value (\ex{5}{7}{} yr; Section 2.2.4) in regions of relatively large $\rho$/T (i.e. the inner disc). The wider spread of the SMC model S1 (i.e. where \tdest{}$>$1 Gyr), is largely attributable to diffuse gas particles in the line-of-sight stripped from the disc plane by tidal forces from the LMC/MW. This interpretation is further supported by the time-varying ratio of \tdest{} for the SMC/LMC (inset panel of Fig. 26), which shows peaks coinciding with pericentres in its complex orbits of the LMC and MW. Nonetheless, our models are implied to have appropriately reproduced the ISM density distribution given the underlying ratio of \tdest{} in the SMC/LMC  at present time ($\sim$3) is consistent with the aforementioned estimates.

%%%%%%%%%%%%%%%%%%%%%%% 4.2
\subsubsection{Depletion in the Stream}

% initial Rdm high, acceptable for an old system like the MCs, so sptial distribution likely key
% FIR emission maps of the MCs indicate that dust peaks in concentration in the inner disc/bar ends, Fig. 21
% dust only forms in cool dense clouds (Draine 2009), which our models concentrate towards the centre
% real MCs show broad turbulent ISM, our models show significant concentration due to tidal infall, indicating further refinement of mass profile/SN feedback

It is not clear from observation whether the Magellanic Stream and Leading Arms harbour significant extant dust mass (i.e. Fong et al. 1987; Br\'{u}ns et al. 2005), in spite of clear depletion patterns similar to the Galaxy (Fox et al. 2013). The extraneous \hi{} streams of our simulated MCs do not follow the universal dust-to-metal relation to which their parent discs are bound (Figure 23), although we note that the tenuous tails test the limits of our Galaxy-derived dust model. 

The lack of direct detection for stardust sources in the Stream presently negates {\it in situ} formation, while accretion is limited by the available refractory elements. Reprocessing of mantle into further core material provides a speculative remedy for dust growth in the Stream. Dwek (1998) invokes this process to explain the failure for the observationally-inferred destruction timescale by SNR in the Galaxy (Jones et al. 1996) to account for core depletion of Iron. Accretion may be further limited by the systematic absence of \htw{} in the observed and simulated Stream (Figure 6). Zwaan \& Prochaska (2006) show for local galaxies a transition from \hi{} to \htw{} at a column density of 10$^{\rm 20}$ \cms, which surpasses the peak observed densities of the Magellanic Stream (Putman et al. 2003; Br\'{u}ns et al. 2005). Our high-resolution simulations are presently insufficient for probing this transition in highly localised molecular clumps within the diffuse Stream. This \htw{} dependence may be irrelevant; Lisenfeld \& Ferrara (1998) find in a sample of dwarf galaxies, strong depletion in spite of the lack of dense molecular clouds in which accretion would occur. Moreover, at up to $\sim$1.7 Gyr old, the tails are not young; extant dust could be long-lived with the relative lack of ionizing stars in the Stream, but we do not incorporate ionization from hot halo interactions that can permeate the densest self-shielding regions of the Stream. 

At relatively low mass, with strong SN activity as evidenced by a porous ISM (i.e. Kim et al. 1998), the MCs are susceptible to the outflow of heavy elements, a mechanism which has been proposed to explain dispersions in \rdm{} for dwarf galaxies (Lisenfeld \& Ferrara 1998). The existence of high velocity clouds on the periphery of, and traced to, the LMC (Staveley-Smith et al. 2003) has also motivated previous studies to suggest the Stream is a product of outflow (Olano 2004; Nidever et al. 2008). This is partly in response to the paucity of stars in the Stream, although our simulations suggest that the tidal \hi{} will not necessarilty have a stellar counterpart. A further refutation arises from abundance measures of the metal-rich Stream component (Fairall 9). Ejecta from SNe type II would show a deficiency in C, N and Fe relative to O (Garnett 2002), but this is not exhibited in the depletion analysis (Figure 8 of Fox et al. 2013). On the other hand, the mixing of dust and exchange with the gas-phase means that the observed depletion may be more a measure of grain survival than the progenitor stellar sources (Dwek 1998).

We posit two interpretations for the depletion level and apparent absence of dust in the Stream, in the context of tidal/ram pressure stripping from the SMC disc. First, the depletion level of the Stream reflects that of the MCs (prior to stripping 2 Gyr ago) which, in our simulations, depends sensitively on SF parameters, feedback, and dust lifetimes (Section 3.1.5). Moreover, dust typically traces metallicity; an appropriate choice for the initial metallicity gradient is as yet ill-defined within observed ranges (Zaritsky et al. 1994); both MCs show flattened profiles at present time, but are likely to exhibited steeper gradients previously (Rubele et al. 2012), before recent enivronmental influences in the Local Group and bar-driven gas inflow became significant. 

Secondly, recent observations and simulations suggest that a group/cluster environment is more efficient at removing \hi{} than \htw{}/dust (Pappalardo et al. 2012). B13 interprets these results in terms of dust being locked within cool stabilised \htw{} regions, as opposed to its relatively loose \hi{} envelope. A corollary of this preferential stripping is a positive radial gradient in the dust-to-gas ratio, which to date has not been reproduced numerically (B13; this work). Other evidence suggests however no such preference. Stripped dust is evident in the SMC tail (Gordon et al. 2009); the corresponding D in these regions is less than expected from metallicity, but the authors propose this as a consequence of local destruction. Roussel et al. (2010) find stripped dust from M82 following a tidal interaction, which they distinguish from that swept out by superwinds, which is warmer and more diffuse. Walter et al. (2007) also find substantial dust mass in the tidal arm of NGC 3077; this feature was pre-enriched before its stripping $\sim$\ex{3}{8}{} yrs ago, given that the SFR within the arm is insufficient to form the dust. A more detailed study of the stripping efficiency of various subcomponents of the ISM will comprise a future paper.

%%%%%%%%%%%%%%%%%%%%%%%%%%%%%%%%%%%%%%%%%%%%%%%%%%%%%%%%%%%%%%%%%%%%%%%%%%%%%%%
\begin{figure}
\includegraphics[width=1.\columnwidth]{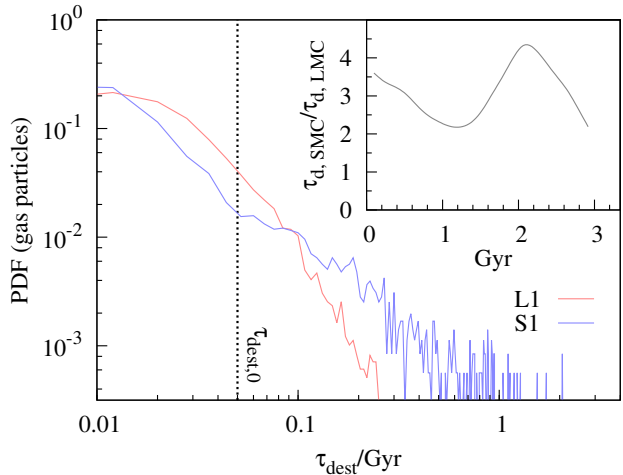}
\caption{Probability Distribution Function (PDF) for gas particles in models L1 and S1 at present time, as a function of their local destruction timescale (\tdest{}, as calculated from Equation 7). The dotted line signifies the reference \tdesto() from the Galaxy; (Inset) The ratio of mass-averaged \tdest() in the S1 and L1, as a function of lookback time.}
\end{figure}

%%%%%%%%%%%%%%%%%%%%%%%
\subsubsection{Polycyclic Aromatic Hydrocarbons}

Up to 20 per cent of the IR discharge of dust contaminated objects show emission lines corresponding to PAHs (Smith et al. 2007), but their origin is not well defined (Paradis et al. 2009). A recent census of PAH-producing carbon stars in the LMC suggests that stellar sources provide a magnitude less than the total PAH mass over their assumed lifetime (Matsuura et al. 2013). The implication is that the ISM reprocesses grains, supported by the anti-correlation of PAHs with carbon-rich stars in the LMCs (Sandstrom et al. 2010) and its preferential residence in dense molecular gas (Wiebe et al. 2011). The production of PAHs from the shattering of larger grains in shocks (Jones et al. 1996) is also proposed from the spatial anti-correlation of PAHs and very small grains (VSGs) at the surfaces of molecular clouds (Paradis et al. 2009). In this scenario, PAHs are longer lived than larger grain species, perhaps explicating their spatial discordance with the original progenitor carbon stars. 

In the first self-consistent models of PAHs, B13 delegated 5 per cent of the total carbonaceous dust in the ejecta of C-rich AGB stars to new PAHs, similar to the abundance fraction in the Milky Way (0.046, Li \& Draine 2001). For the LMC, this model is reasonable given the majority of carbon-rich stars lie in the vicinity of the stellar bar (Matsuura et al. 2009), coincident with a PAH mass enhancement (Paradis et al. 2009; Meixner et al. 2010). B13 postulates that this enhancement stems from tidal interactions triggering centralised starbursts; the environment of the LMC could also explain its greater PAH abundance over galaxies of comparable metallicity (Paradis et al. 2009). 

In vindication of this model, B13 finds the PAH-to-dust mass fraction varies with metallicity in accord with observations (Draine et al. 2007; Mu\'{n}oz-Mateos et al. 2009). In a cursory diagnosis of our reference MC models, we find the PAH-to-dust mass ratio overlaps with the observed $\sim$1-2 per cent in star forming regions, to 0.6 per cent in diffuse ISM and \hii{} regions (Sandstrom et al. 2011). We note however that our models show positive PAH abundance gradients, in conflict with observed samples of Sdm and Irr galaxies (Mu\'{n}oz-Mateos et al. 2009) which show a preference for zero to negative gradients. This partly stems from the non-inclusion of old C-rich AGBs (age greater than our simulation duration of 3.4 Gyr) that would have settled towards the galaxy centre. Nonetheless, PAH destruction is likely the driving factor behind the apparent metallicity dependence (Bolatto et al. 2007). Observed spatial variations in PAH emission strength, in particular those around 30 Doradus, suggest that PAHs tend to avoid bright H$_{\rm II}$ regions and are found instead near atomic and molecular gas (Wiebe et al. 2011). We thus advocate future models of the MCs to model the various evolutionary paths of PAHs to further shed light on dust lifecycles in the MC ISM.

%%%%%%%%%%%%%%%%%%%%%%%%%%%%%%%%%%%%%%%%%%%%%%%%%%%%%%%%%%%%%%%%%%%%%%%%%%%%%%%
\subsection{Molecular Hydrogen in the MCs}

%%%%%%%%%%%%%%%%%%%%%%%%%%%%%%%%%%%%%%%%%%%%%%%%%%%%%%%%%%%%%%%%%%%%%%%%%%%%%%%
\subsubsection{Abundance as a function of ambient conditions}

The global \htw{} mass fraction of our simulated LMC is consistent with observational estimates ($\sim$10 per cent; Israel 1997; Fukui et al. 1999), is comparable to wider samples of such galaxies (Young \& Scoville 1991), and accordingly less than the Galaxy. The simulated SMC, under similar evolutionary parameters, produces an \htw{} abundance of up to 15 percent. This lies in excess of observational estimates (again, $\sim$10 percent), including those that account for a CO-dark \htw{} component for $A_{\rm O}<8.2-8.4$ (Leroy et al. 2007; Wolfire et al. 2010; Leroy et al. 2011), in spite of a broad agreement between observations for the \hi{} mass/metallicites of the MCs and our models. 

Since \htw{} formation is sensitively bound to the dust abundance, this excess can be traced to oversimplifications in our adopted dust production model (Section 4.1.1). Furthermore, the flux density from irradiating massive stars is fixed to SEDs derived from the Salpeter IMF (Section 2.2.7), ignoring the possibility for a top-heavy IMF for lower metallicity galaxies (Maio et al. 2007). Refinements in this direction would permit comparison with the varying hardness of the UV field (Cox et al. 2006), especially that between individual clouds (Paradis et al. 2011). Greater regulation by photo-ionisation would manifest in an increasing average cloud column density with decreasing metallicity to attain equilibrium (McKee 1989). The ambient UV radiation is recognised as generally harsher than the MW, and clouds are accordingly hotter and smaller in the SMC (Lequeux et al. 1994; Leroy et al. 2007).

%%%%%%%%%%%%%%%%%%%%%%%%%%%%%%%%%%%%%%%%%%%%%%%%%%%%%%%%%%%%%%%%%%%%%%%%%%%%%%%
\subsubsection{Distribution as a function of dynamics}

Recent studies have suggested the metallicity-dependent balance between formation and dissociation is not as critical as ISM turbulence (Walch et al. 2011; Renaud et al. 2012) and local density enhancements (Mac Low \& Glover 2012). For example, GMCs and YSO in the LMC are often enhanced and organized in the vicinity of supergiant shells and their colliding fronts (Dawson et al. 2012). The turbulent ISM is also driven by successive tidal interactions that render the ISM very porous; several hundred supershells in the SMC (Staveley-Smith et al. 1997) conform to a turbulence spectrum driven by the recent close encounter (Goldman 2000).

Detailed simulations of the MCs (Bekki \& Chiba 2007; Besla et al. 2012; this work) cannot yet accurately reproduce the present distribution of star formation or \htw{}; a clear example in our models is the concentration of \htw{} in the stellar bars, in a manner similar to massive barred spirals, but unlike the scattered FIR/CO emission in both MCs (i.e. Wong et al. 2009). Moreover, numerical limtations restrict our capacity to address the possible preferential stripping of \htw{} over \hi{} (Pappalardo et al. 2012); the stronger tidal actions acting on the lower mass SMC may explain why its \htw{} gas mass fraction is broadly similar to that of the LMC in spite of a factor $\sim$2.5 lower mean metallicity.

On the other hand, Figure 19 shows an analogue to the \hi{} overdensity (Nidever et al. 2008) or Molecular Ridge, which at 100-200 pc wide and 1.8 kpc long (Ott et al. 2008) is argued to be a coherent structure, but kinematics and the discrepancy from the \hi{} envelope would suggest otherwise. Extending from one bar end, without a clear counterpart at the other, the structure forms in our simulations from strong recent lopsidedness (Section 3.3.2) of the bar resulting in the asymmetric collision of gas flows. The hydrodynamical collision of the LMC and SMC or hot halo have also been proposed for its formation (Fujimoto \& Noguchi 1990; Kim et al. 1998), but which are not modelled here. 

Significant molecular mass is also traced in the simulated Bridge (Section 3.4); the Stream, at comparable metallicity, is conversely devoid of \htw{} (Figure 6). Young stars with ages 10 to 40 Myr are located throughout the Bridge (Demers \& Battinelli 1998), including young B-type stars with a low (0.1 \zd{}) metallicity similar to the gas phase ([Fe/H]$=$-1.7 to -0.9; Lehner et al. 2008). The low metals abundance clearly does not preclude \htw{} formation and subsequent star formation in this \hi{}-rich environment. In a similar bridge structure between massive spirals, Gao et al. (2004) find the highest concentration of \ha{} near the {\it intruder} galaxy (which in our simulations is the SMC), due to the collisional nature of the gas-rich interaction.

%%%%%%%%%%%%%%%%%%%%%%% 4.4
\subsection{Other influences on the Age Metallicity Relation}

We propose that the observed AMR is a complex function of the role played by dust evolution in quiescent and starburst epochs; the modest but sustained enrichment of the MCs since 3.4 Gyr ago (Weisz et al. 2013) disguises an underlying balance of competing growth and destructive processes in a tidally-driven ISM, given the coupling of dust mass/ metallicity. In the following sections, we suggest refinements to our models and model analysis that would further elucidate the evolution of the MCs.

%%%%%%%%%%%%%%%%%%%%%%%%%%%%%%%%%%%%%%%%%%%%%%%%%%%%%%%%%%%%%%%%%%%%%%%%%%%%%%%
\subsubsection{Mass transfer and infall}

The tidally-driven MCs would show evidence in their AMR of not only the nominally secular balance of dust accretion and destruction, but also the infall and outflow of ISM. Tidal torques retard gas in the disc such that it encourages infall and fuels nuclear starburst and centrally concentrated growth, with the central starburst determined by the infall rate (Jog \& Das 1992). In one-zone models, dispersions in the infall rate have been shown to perturb the derived \rdm{} (Inoue 2003), while the metallicity drops if the rate of gas accretion exceeds the SFR (Koppen \& Edmunds 1999). 

The recent interaction history of the MCs provides multiple means of gas infall onto the galaxies with the implication of dilution, but most studies associate recent close encounters with rapid enrichment (Harris \& Zaritsky 2009; Livanou et al. 2013). This mirrors a broader debate on the role of interactions on metallicity: Marquez et al. (2002) find in a large sample of isolated and interacting spirals that the gas-phase metallicity does not appear to correlate with the interaction state, while conversely Kewley et al. (2006) find smaller galaxy separations correlated with lower metallicity. 

Evidence for major infall events in the MCs lie first with the assertion by Bekki \& Tsumjimoto (2012) that an apparent dip in the AMR (Harris \& Zaritsky 2009) corresponds to stripped metal-poor gas from a less massive SMC, consistent with a subpopulation of anomalous stars in the LMC which Olsen et al. (2011) propose originate from the SMC. The mass transfer invoked by Bekki \& Tsujimoto (2012) to drive this dip exceeds that simulated in this work. Alternatively, the timescale of this dip agrees with simulations of fly-by interactions (Montuori et al. 2010), wherein inflows occur over a dynamical time ($\sim$10$^{\rm 8}$ yr) and the dilution of circumnuclear metallicities persist up to 10$^{\rm 9}$ yrs; the extent of the dip is also quantitatively similar to the 0.1 to 0.3 dex dilution established by Rupke et al. (2010). For the more recent close interaction ($\sim$0.2 Gyr ago), Bekki \& Chiba (2007) show that 10$^{\rm 8}$ \md{} of gas can be stripped from the SMC and collide with the LMC. The relative velocity of gas transferred in this scenario ($\sim$60 kms$^{\rm -1}$) is less than that of HVCs ($\sim$160 kms$^{\rm -1}$) and the circular velocity of the LMC disc ($\sim$80-120 kms$^{\rm -1}$), and would better explain the young ($\sim$10 Myr) subpopulation of Nitrogen deficient stars than the timescales of accreting HVCs.

Gas accretion from the IGM is another efficient driver of secular evolution (Dekel et al. 2009). The stellar bar of the LMC is barely traced by the ISM (Mizuno et al. 2001), suggesting it imparts a weak gravitational influence. The lifetime of bars with a dissipative medium is not well constrained, but weakening can occur gradually by cold gas accretion (Berentzen et al. 2004). Concurrently, accretion can enhance asymmetry in strength and longevity (Bournard et al. 2005). The inner LMC and bar show strong lopsidedness (van Der Marel et al. 2002), the persistence of which is implied by carbon-rich stars which preferentially lie at one bar end for at least a Gyr (Cioni, Habing \& Israel 2000; Matsuura et al. 2009). Bekki \& Tsujimoto (2010) hypothesise that accreted HVCs are responsible for the low nitrogen content of \hii{} regions (a factor of 7 lower than solar), which mix within a rotation time (10$^{\rm 8}$ yr), and have minimal impact on other elemental abundances. The implications of a gas supply external to the MCs has yet to be modelled self-consistently and is advocated for future studies.

%%%%%%%%%%%%%%%%%%%%%%%%%%%%%%%%%%%%%%%%%%%%%%%%%%%%%%%%%%%%%%%%%%%%%%%%%%%%%%%
\subsubsection{Stellar outflow}

The metallicity and kinematics of HVCs on the periphery of the LMC imply their origin from the gas disc, following superwind driven outflow (Staveley-Smith et al. 2003), which can also explain the various components of \hi{} in the line-of-sight of the MCs. The metal-rich filament lying ahead of Fairall 9 (Richter et al. 2013) also advocates the LMC as an origin to the Stream (Olano 2004; Nidever et al. 2010), although the elemental abundance may negate this argument (Section 4.1.4). Alternatively, the stronger outflow from the lower mass SMC could strip metal-rich ISM from LMC to form this filament, a mechanism explored for generic low and intermediate mass galaxies by Scannapieco \& Broadhurst (2001).

These arguments are consistent with the prominent role of SNe in gas-rich, low mass galaxies with inefficient star formation akin to the SMC (Dalcanton 2007). Lisenfeld \& Ferrara (1998) attribute the wide dispersion in \rdm{} to outflow, where metals mixed in hot gas preferentially escape, while the cool ISM is largely retained (MacLow \& Ferrara 1999). The mass range susceptible to metal-enriched outflow ($<$10$^{\rm 10.5}$ \md{}; Garnett 2002) includes our models for the SMC and low mass LMC. 

Strong feedback is also invoked to sustain the late-type spiral structure of the LMC, by mitigating the quenching, gas depletion and build-up of a nuclear bulge that follows strong starburst phases (Governato et al. 2009). This process shares a complex relationship with accretion mechanisms discussed in Section 4.3.2, where feedback drives out gas and mitigates subsequent infall through the dynamical heating of the halo (Dubois \& Teyssier 2008). The multiple recent starbursts of the MCs can also mitigate outflow, where dynamical heating by previous feedback activity can oppose the motion of more recent superbubbles. 

The relationship between infall and outflow thus represents an important question for future simulations of Magellanic-type galaxies, where presently adopted models are too primitive for this purpose. Future work could follow from previous one-zone model considerations of [$\alpha$/Fe] as a means of constraining starburst epochs and gas outflow (Bekki \& Tsujimoto 2012), who find variations from a Salpeter IMF, or outflows with preferential [$\alpha$/Fe] can explain elemental abundances. 

%%%%%%%%%%%%%%%%%%%%%%%%%%%%%%%%%%%%%%%%%%%%%%%%%%%%%%%%%%%%%%%%%%%%%%%%%%%%%%%
\section{Conclusion}

We have investigated the recent chemodynamical evolution of the Magellanic System with N-body/SPH simulations, and the first self-consistent model of dust and \htw{} lifecycles. Our models are broadly consistent with observations; we discuss these results in terms of the metallicity-dependence of dust production, dynamical influences on \htw{}-abundance, and future refinements to the numerical model. The principal results are summarised as follows:

\begin{enumerate}

\item 
We simulate the Magellanic Clouds for their previous 3.4 Gyr of chemodynamical evolution while in proximity to the Galaxy, based on collisionless models originally developed in Diaz \& Bekki (2012). This timescale corresponds to a period of elevated star formation (Weisz et al. 2013), which our models associate with strong mutual tidal perturbations operating within the LMC-SMC-Galaxy triplet. The past orbits of the MCs are reconstructed from recent proper motion measurements and constrained by the morphology of the Magellanic Stream, which is formed from tidal stripping of the SMC commencing $\sim$2 Gyr ago. The mass composition of our SMC model S1 (which we assume includes most of the neutral hydrogen that presently comprises the Magellanic Stream, Bridge and Leading Arms) and LMC model L2 are thus constrained by that adopted in Diaz \& Bekki (2012) wherein observed rotation curves were reproduced. We also investigate the evolution of a massive LMC model L1 whose mass more closely matches that predicted by pre-infall halo occupation models.

\item
Our ISM model, which explicitly follows the evolution of \htw{} and multi-elemental dust, incorporates parametrizations derived in the seminal work of Dwek (1998) and similarly assumes the 2:1 coupling of dust destruction/accretion timescales. We calibrate this model, together with star formation criteria, against the observed star formation law and dust-to-metal abundance relations of the Clouds. Our reference model for the SMC reproduces the present star formation rates inferred from FIR and \ha{} emission (i.e. Bolatto et al. 2011). We find qualitative agreement with the observed age-metallicity relations and star formation histories of both MCs (Harris \& Zaritksy 2004; 2009; Weisz et al. 2013). For low and high mass LMC models, a recent ($<$300 Myr ago) enhancement in SFH coincides with above-average asymmetry of the disc as observed (van der Marel et al. 2002) and representative of the wider class of Magellanic Irregulars.  

\item 
In a parameter study, we find the overall enrichment of the metal-poor Clouds highly dependent on the choice of dust parameters, in particular the characteristic timescale for destruction by neighbouring SNe. Assuming destruction is most efficient in supernova remnants and is thus correlated with the recent star formation rate, the ratio of mean destruction timescales in our LMC and SMC models is consistent with other observational and analytical studies.

\item 
By adopting an initial dust-to-metal ratio of 0.4 (the presumed metallicity-independent value for evolved systems) our reference LMC/SMC models match the observed ISM abundances. Moreover, the models are bound to this ratio throughout their evolution. Commencing from a ratio of 0.1 converges upon the same abundances, implying the tendency towards an equilibrium state. Observational estimates of the dust-to-gas ratio and oxygen abundance of the MCs and other metal-poor systems favour instead a non-linear metallicity dependence of dust abundance. We trace this discrepancy to the uncertainty regarding SNe II dust condensation efficiencies; our adopted values from Dwek (1998) are ostensibly too high, given recent evidence suggesting that dust growth in the ISM is the primary source of the present dust mass.

\item
The morphology and low (0.1 \zd{}) metallicity of the simulated Magellanic Stream and Leading Arms (i.e. Fox et al. 2013) are reproduced following the tidal stripping of our SMC model by the MW/LMC. Our simulations are the first to model these tidal features with an ISM comprising \hi{}, \htw{} and dust components; we find they are dominated by \hi{}, while \htw{} and dust make a negligible contribution. The \hi{} shows a column density gradient along the Stream, similar to that observed (Putman et al. 2003) and thus precludes the requirement for ram pressure interactions with the Galaxy to explain this feature. The \hi{} mass is however deficient by a factor of $\sim$5, similar to that found by other numerical sutides (i.e. Besla et al. 2012). This highlights a fundamental underestimation of the SMCs original mass/gas budget.

\item
The dust-to-gas ratio and depletion of the gas phase to dust condensate is smaller in our simulated Stream than that implied from sightline analysis, attributable to the stripped material being loose, unenriched metal/dust-poor gas from the peripehery of the SMC. We show in additional models how the observed $\sim$0.6 dex depletion level can be obtained from encouraging dust accretion in the SMC gas disc prior to its stripping to the Stream; these models lead however to excessive \htw{} and dust abundances in the parent disc. We also address the recent hypothesis that a metal-rich and kinematically distinct filament in the Stream was sourced from the LMC (Nidever et al. 2010; Richter et al. 2013). Unlike our massive LMC model (dynamical mass \ex{6}{10}{} \md{}), our low mass model (\ex{1}{10}{} \md) contributes a filament to the Stream via stripping by the MW. This filament is less massive than that sourced from the SMC by a magnitude, and does not readily match the metal or dust abundances of the aforementioned filament for the same reason as the SMC-stream. 

\item 
In a subsequent paper, we aim to improve upon our reproduction of the MCs evolution and present abundances (in a tidal-dominated scenario with the adopted dust/\htw{} model) with the following major refinements: 1) uncoupling the relationship between dust accretion and destruction timescales to better reflect local ISM conditions; 2) calibrating our SNe II condensation efficiencies and characteristic timescales of dust growth against those estimated for metal-poor systems; and 3) extending the method of Diaz \& Bekki (2012) to larger lookback times (assuming improved constraints on the MCs proper motions and dynamical masses become available), such that we can assert if a realistic ISM abundance distribution at the onset of recent tidal stripping can accomodate the observed depletion of the Stream.
 
\end{enumerate}

%%%%%%%%%%%%%%%%%%%%%%%%%%%%%%%%%%%%%%%%%%%%%%%%%%%%%%%%%%%%%%%%%%%%%%%%%%%%%%%
\section*{Acknowledgements}
We are grateful to the referee for their useful and insightful comments. CY is supported by the Australian Postgraduate Award Scholarship. This research was supported by resources awarded under Astronony Australia Ltds ASTAC scheme at Swinburne with support from the Australian Government. gSTAR is funded by Swinburne and the Australian Governments Education Investment Fund. This work was also supported by iVEC through the use of advanced computing resources located at iVEC@UWA.

\newcommand{\bib}[1]{\bibitem[\protect\citeauthoryear{}{}]{} #1}

%references 213

%\appendix 

\bsp
\label{lastpage}
\end{document}